# Low-frequency unsteadiness mechanisms in shock wave/turbulent boundary layer interactions over a backward-facing step

### Weibo Hu†, Stefan Hickel and Bas W. van Oudheusden

Faculty of Aerospace Engineering, Delft University of Technology
Kluyverweg 1, 2629HS, Delft, The Netherlands

The low-frequency unsteady motions behind a backward-facing step (BFS) in a turbulent flow at $Ma = 1.7$ and $Re_\infty = 1.3718 \times 10^5$ is investigated using a well-resolved large-eddy simulation (LES). The instantaneous flow field illustrates the unsteady phenomena of the shock wave/boundary layer interaction (SWBLI) system, including vortex shedding in the shear layer, the flapping motions of the shock and breathing of the separation bubble, streamwise streaks near the wall and arc-shaped vortices in the turbulent boundary layer downstream of the separation bubble. A spectral analysis reveals that the low-frequency behaviour of the system is related to the interaction between shock wave and separated shear layer, while the medium-frequency motions are associated with the shedding of shear layer vortices. Using a three-dimensional dynamic mode decomposition (DMD), we analyse the individual contributions of selected modes to the unsteadiness of the shock and streamwise-elongated vortices around the reattachment region. Görtler-like vortices, which are induced by the centrifugal forces originating from the strong curvature of the streamlines in the reattachment region, are strongly correlated with the low-frequency unsteadiness in the current BFS case. Our DMD analysis and the comparison with an identical but laminar case provide evidence that these unsteady Görtler-like vortices are affected by fluctuations in the incoming boundary layer. Compared to SWBLI in flat plate and ramp configurations, we observe a slightly higher non-dimensional frequency (based on the separation length) of the low-frequency mode.

## 1. Introduction

Shock wave/boundary layer interaction (SWBLI) has been an active research topic in the aerospace community over the past decades. This flow phenomenon is ubiquitous in high-speed aerodynamics, such as supersonic inlets, over-expanded nozzles, high-speed aerofoils (Green 1970; Dolling 2001). Shock-induced boundary-layer separation is a main contributor to flight drag of transonic aerofoils and pressure loss in engine inlets, which illustrates its relevance. Moreover, significant fluctuations of pressure and temperature are widely observed around the interaction regions. SWBLI can cause intense localized mechanical and thermal loads, which may eventually lead to the failure of material and structural integrity (Délery & Dussauge 2009; Gaitonde 2015). It is therefore crucial to take the effects of SWBLI into account in the process of aircraft design and maintenance, including material selection, assessment of fatigue life and thermal protection systems.

Canonical two-dimensional SWBLI configurations can be abstracted into three simplified cases: (1) incident (impinging-reflecting) shock, (2) compression ramp and (3) backward/forward-facing step (BFS/FFS). Considerable progress has been achieved in understanding the unsteady phenomena and underlying mechanisms of SWBLI by means of advanced flow measurement techniques and well-resolved numerical simulations, particularly for the flat plate impinging shock and compression ramp configurations (Ganapathisubramani *et al.* 2007; Grilli *et al.* 2013; Pasquariello *et al.* 2017). These two cases share similar mean flow topology although the shocks are produced by different mechanisms, as shown in Figure 1(a) and (b). In the impinging/reflecting shock case, the incident shock induces a strong adverse pressure gradient on the boundary layer, which leads to the separation of the boundary layer. A separation shock is produced ahead of the separation point and a reattachment shock is generated around the reattachment location due to the compression of the boundary layer. For the ramp case, the strong flow compression caused by the ramp geometry induces a strong (separation) shock, which results in the separation of the incoming boundary layer. Subsequently, a reattachment shock is generated as the separated shear layer reattaches on the ramp downstream. In both cases, the SWBLI is accompanied by energetic unsteady motions at frequencies that are one or two orders lower than the boundary layer characteristic frequency $u_\infty/\delta$ (Touber & Sandham 2009, 2011). Considerable research effort has been put into tracing the source of this low-frequency unsteadiness.

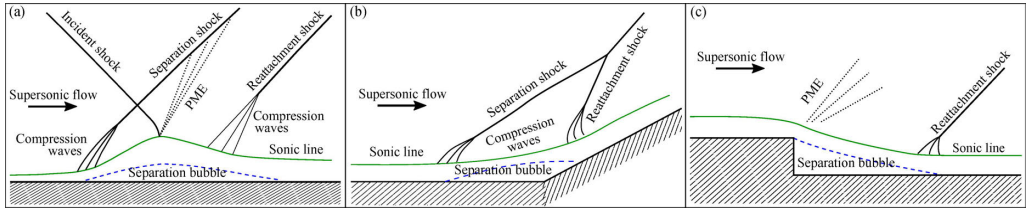

Figure 1: Mean flow structures of SWBLI in canonical two-dimensional configurations (a) impinging shock, (b) compression ramp and (c) backward-facing step.

In general, theories regarding the origin of this low-frequency motion of the separation shock are categorized as resulting from either upstream or downstream dynamics. The first group of theories associates the unsteady motions with upstream fluctuations within the incoming turbulent boundary layer. In an early work, Plotkin (1975) proposed a simple linear restoring model to explain the source of the shock wave oscillations, in which the shock is displaced by velocity fluctuations inside the upstream turbulent boundary and tends to return to its mean location through a restoring mechanism determined by the stability of the mean flow. The pressure measurement by Andreopoulos & Muck (1987) provided the first experimental evidence for a correlation of the shock wave unsteadiness with bursting events upstream the boundary layer in a compression ramp case at $Ma = 1.7$. Unalmis & Dolling (1996) found low-frequency pressure fluctuations along the spanwise direction in the incoming boundary layer by measuring the pressure signal in the ramp case at $Ma = 5$. Poggie & Smits (2001) performed measurements of wall pressure fluctuations and schlieren visualization in a backward-facing step/ramp configuration at $Ma = 2.9$. They reported that also in this case the shock motion was correlated with upstream large-scale wave structures. Based on the cross-correlation analysis, they concluded that their experimental results are in good agreement with the linear restoring mechanisms proposed by Plotkin (1975). Beresh *et al.* (2002) used particle image velocimetry (PIV) and high-frequency response wall pressure transducers for a compression ramp interaction, and they found a clear correlation between streamwise velocity fluctuations in the lower part of the upstream boundary layer and low-frequency shock motions. In addition, they found no correlation between shock oscillations and the velocity fluctuations in the upper part of the upstream boundary layer, as well as the variation of the upstream boundary layer thickness, as reported by McClure (1992) in earlier work. Ganapathisubramani *et al.* (2007) also observed elongated superstructures with low- and high-speed streaks upstream the separation region in their stereoscopic PIV and planar laser scattering (PLS) measurements of a Mach 2 compression ramp interaction and they proposed these upstream large-scale structures are responsible for the low-frequency unsteadiness of the interaction region. Humble *et al.* (2009)

further confirmed the presence of streamwise-elongated low- and high-speed streaks inside the upstream boundary layer using tomographic PIV for an incident shock interaction at $Ma = 2.1$. Their results show that this reorganization of the upstream boundary layer in both streamwise and spanwise directions conforms to the overall streamwise translation and spanwise rippling of the interaction region. However, Touber & Sandham (2011) argued that the low-frequency interaction motions do not necessarily require a forcing source from upstream or downstream and are more like an intrinsic response to the broadband frequency spectrum of the upstream turbulent fluctuations. Porter & Poggie (2019) consider that this response is a selective response of the separation region to certain large-scale perturbations in the lower half part of the upstream boundary layer based on their high-fidelity simulation.

The second group of theories attributes the low-frequency dynamics to mechanisms intrinsic to the interaction system itself, that is, with an origin downstream of the separation line. Already early experimental studies suggested that the low-frequency motion of the separation shock is linked to the expansion and contraction of the separation bubble (Erengil & Dolling 1991; Thomas *et al.* 1994). For the impinging shock induced interaction, Dupont *et al.* (2006) found a clear statistical link between low-frequency oscillation of the separation shock and the downstream interaction region by analysing experimental pressure signals. Furthermore, they also reported a quasi-linear relation between the separation shock and the reattachment shock motions. By DNS of a Mach 2.25 impinging shock case, Pirozzoli & Grasso (2006) established a resonance theory, in which acoustic waves are produced by the interaction between coherent structures in the bubble and the incident shock. The upstream propagation of these acoustic waves is responsible for the low-frequency oscillations of the SWBLI system. Touber & Sandham (2009) performed a global linear stability analysis of the mean flow field from their LES and detected an unstable global mode inside the separation bubble, which provides a possible driving mechanism for the low-frequency unsteadiness by displacing the separation and reattachment points. Piponniau *et al.* (2009) proposed a simple physical model that relates the low-frequency oscillations to the breathing motions of the separation bubble, in which the collapse of the separation bubble is caused by a continuous entrainment of mass flux, while the dilation corresponds to a radical expulsion of the mass injection in the bubble. A similar model was suggested by Wu & Martín (2008) based on DNS of a compression ramp configuration. They consider that a feedback loop, involving the separation bubble, the detached shear layer and the shock system, is the underlying mechanism for low-frequency shock motions. The DMD analysis of Grilli *et al.* (2012) provided further evidence that mixing across the separated shear layer leading to a contraction and expansion of the separation bubble is the dominant mechanism for the low-frequency unsteadiness. Numerical work of Grilli *et al.* (2013) and Priebe *et al.* (2016) identified streamwise-elongated Görtler vortices originating around the reattachment location for compression ramp configurations. For an impinging shock configuration, Pasquariello *et al.* (2017) reported very similar observations of low-frequency DMD modes characterised by streamwise-elongated regions of low and high momentum that are induced through Görtler-like vortices. As the separation-bubble dynamics is clearly coupled to these vortices, Görtler-like vortices might act as a source for continuous (coherent) forcing of the separation-shock-system dynamics.

In an attempt to resolve this discrepancy, Souverein *et al.* (2010) proposed that actually both upstream and downstream mechanisms contribute to the SWBLI dynamics with case dependent intensity. Which type of mechanism is more dominant in producing the low-frequency dynamics depends on the shock strength and possibly the Reynolds number. In weak interactions, the low-frequency unsteady motions can be mainly associated with upstream effects, while the unsteadiness of the strong interactions are more likely driven by the dynamics of the downstream separation bubble and reattachment shock (Clemens & Narayanaswamy 2014). Also Priebe *et al.* (2016) implied that upstream disturbances contribute to the low-frequency behaviour although they consider that the downstream Görtler instability is the dominant one. Bonne *et al.*

(2019) indicated that the low-frequency oscillations involve both the amplification of upstream disturbances by the separated shear layer and a feedback excitation from the shock foot and backward travelling density waves.

As discussed above, SWBLI in the impinging shock and compression ramp configuration share similar unsteady behaviour and physical mechanisms (Clemens & Narayanaswamy 2014; Smits & Dussauge 2006). In contrast to these well-analysed canonical cases, supersonic flow over a BFS has a distinctly different flow topology, as shown in Figure 1(c). The incoming turbulent flow undergoes first a centred Prandtl-Meyer expansion (PME) with the separation location fixed at the step convex corner. The free shear layer then develops towards the downstream wall on which the flow reattaches. Compression waves are generated around the reattachment location, which coalesce into a reattachment shock (Loth *et al.* 1992; Sriram & Chakraborty 2011). In this configuration, the upstream limit of the separation bubble is stationary and only the downstream reattachment shock is present. The dynamics of the recirculation and shock region is reported to be unsteady as in other conventional cases (Bolgar *et al.* 2018). In an early experimental study, by examining the variation of skin friction, Ginoux (1971) observed the systematic development of counter-rotating streamwise vortices around the reattachment, occurring in laminar, transitional and turbulent flows alike. The wavelength of these vortices is equal to two or three times the boundary layer thickness for a wide range of Mach numbers. These Görtler-like vortices were also reported in the experimental visualization via nano-tracer-based planar laser scattering (NPLS) (Zhu *et al.* 2015). In addition, small unsteady shedding vortices along the shear layer were identified by Chen *et al.* (2012) using the same visualization techniques. However, the Kelvin–Helmholtz (K-H) vortices typical in laminar and transitional cases were not observable in the turbulent shear layer (Zhi *et al.* 2014). The observed coherent vortical structures cover a wide range of length and frequency scales, involving the vortex shedding close to the step, longitudinal vortices and hairpin vortices downstream of the shear layer (Soni *et al.* 2017). The unsteady characteristics can be quantified by the dimensionless Strouhal number $St_r = fL_r/u_\infty$ based on the reattachment length and free stream velocity. By means of particle image velocimetry and dynamic pressure measurements, Bolgar *et al.* (2018) inferred that for a flow at $Ma = 2.0$ the higher frequency content ($St_r = 0.05 - 0.2$) is related to the shock motions, while the dominant low-frequency parts ($St_r \approx 0.03$) are associated with the separation bubble. More efforts are required to scrutinize the frequency characteristics of BFS SWBLI and to analyse whether the low-frequency unsteadiness of supersonic BFS flows has a similar origin as that in the impinging shock and ramp SWBLI cases. In our previous work (Hu *et al.* 2019, 2020), we examined the unsteady SWBLI over a BFS in a laminar inflow regime. The preceding discussion motivates to investigate to what extent the laminar and turbulent cases share similar unsteady features and physical mechanisms.

In this paper, we analyse new large-eddy simulation (LES) results for a fully turbulent BFS flow at $Ma = 1.7$ with special attention to the low-frequency dynamics. The organization of the paper is as follows. Details of the numerical methods used and the setup of the flow configuration are given in §2. Then the flow topology of the mean and instantaneous flow is discussed in §3. The characteristic frequencies of the significant unsteady motions are analysed using spectral analysis. Finally, dominant modes in the SWBLI are extracted via a three-dimensional dynamic mode decomposition (DMD). By comparing with previous works, a physical mechanism of the low-frequency unsteadiness source is proposed (§4). The conclusions with a summary of the main results are presented in §5.

## 2. Flow configuration and numerical setup

### 2.1. *Governing equations*

The physical problem is governed by the unsteady three-dimensional compressible Navier-Stokes equations with appropriate boundary and initial conditions, and the constitutive relations for an ideal gas. We solve the conservation equations for mass, momentum and total energy

$$\frac{\partial \rho}{\partial t} + \frac{\partial}{\partial x_i} \left( \rho u_i \right) = 0 \,, \tag{2.1}$$

$$\frac{\partial \rho u_j}{\partial t} + \frac{\partial}{\partial x_i} \left( \rho u_i u_j + \delta_{ij} p - \tau_{ij} \right) = 0 \,, \tag{2.2}$$

$$\frac{\partial E}{\partial t} + \frac{\partial}{\partial x_i} \left( u_i E + u_i p - u_j \tau_{ij} + q_i \right) = 0 \,, \tag{2.3}$$

where $\rho$ is the density, $p$ the pressure and $u_i$ the velocity vector.

The total energy $E$ is defined as

$$E = \frac{p}{\gamma - 1} + \frac{1}{2} \rho u_i u_i \,, \tag{2.4}$$

the viscous stress tensor $\tau_{ij}$ follows the Stokes hypothesis for a Newtonian fluid

$$\tau_{ij} = \mu \left( \frac{\partial u_i}{\partial x_j} + \frac{\partial u_j}{\partial x_i} - \frac{2}{3} \delta_{ij} \frac{\partial u_k}{\partial x_k} \right) \,, \tag{2.5}$$

and the heat flux $q_i$ is given by the Fourier's law

$$q_i = -\kappa \partial T / \partial x_i \,. \tag{2.6}$$

The fluid is assumed to behave as a perfect gas with a specific heat ratio $\gamma = 1.4$ and a specific gas constant $R = 287.05 \, \mathrm{J(kg \cdot K)^{-1}}$, following the ideal-gas equation of state

$$p = \rho R T \,. \tag{2.7}$$

The dynamic viscosity $\mu$ and thermal conductivity $\kappa$ are a function of the static temperature $T$ and are modelled according to Sutherland's law and the assumption of a constant Prandtl number $Pr$

$$\mu = \mu_{ref} \frac{T_{ref} + S}{T + S} \left( \frac{T}{T_{ref}} \right)^{1.5} \,, \tag{2.8}$$

$$\kappa = \frac{\gamma R}{(\gamma - 1) Pr} \mu \,. \tag{2.9}$$

The values adopted for the computations are: $\mu_{ref} = 18.21 \times 10^{-6} \, \mathrm{Pa \cdot s}$, $T_{ref} = 293.15 \, \mathrm{K}$, $S = 110.4 \, \mathrm{K}$ and $Pr = 0.72$.

### 2.2. *Flow configuration*

The current computational case is an open BFS (i.e., no upper wall) with a supersonic turbulent boundary layer inflow, a schematic of which is shown in Figure 2. The origin of the Cartesian coordinate system is placed at the step corner. The turbulent inflow is characterised by the freestream Mach number $Ma_\infty = 1.7$ and the Reynolds number $Re_{\delta_0} = 13718$ based on the inlet boundary layer thickness $\delta_0$ ($99\% u_\infty$) and free-stream viscosity. The main flow parameters are summarized in Table 1, where we indicate freestream flow parameters with subscript $\infty$ and stagnation parameters with subscript 0. The size of the computational domain corresponds to $[L_x, \, L_y, \, L_z] = [110\delta_0, \, 33\delta_0, \, 16\delta_0]$ including a length of $40\,\delta_0$ upstream of the step in order to exclude potential uncertain effects from the numerical inlet boundary conditions on the flow in the region of interest. The height of the step $h = 3\delta_0$ is three times larger than the inlet boundary layer thickness. In addition to this (fully) turbulent BFS flow, we also present selected results for a fully laminar inflow case with the same free stream flow parameters and geometry (Hu *et al.* 2019) for comparison. Note that this laminar inflow case is referred to as the laminar case for the simplicity of the discussion although flow transition to turbulent occurs shortly downstream of the step.

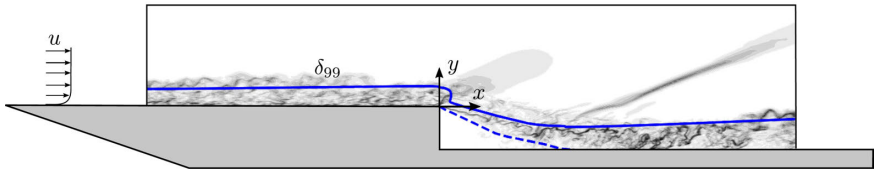

Figure 2: Schematic of the region of interest, which is in the center of the computational domain with the size of $([-40, 70] \times [-3, 30] \times [-8, 8])\delta_0$ in the $x$, $y$, $z$ directions. The figure represents a typical instantaneous numerical schlieren graph in the x-y cross section. The blue dashed and solid lines signify isolines of $u = 0$ and $u/u_e = 0.99$ from the mean flow field.

| $Ma_\infty$ | $U_\infty$ | $\delta_0$ | $\theta_0$ | $Re_\infty$ | $T_0$ | $p_0$ | $h$ | $p_\infty$ |
|---|---|---|---|---|---|---|---|---|
| 1.7 | 469.85 m/s | 1 mm | 0.11 mm | $1.3718 \times 10^7$ m$^{-1}$ | 300 K | $1 \times 10^5$ Pa | 3 mm | 20 259 Pa |

Table 1: Main flow of the current case

### 2.3. *Numerical method*

The LES method of Hickel *et al.* (2014) is used to solve the governing equations. The sub-grid scale model is fully merged into a non-linear finite-volume scheme provided by the adaptive local description method (ALDM) (Hickel *et al.* 2006, 2014). ALDM is based on a solution-adaptive reconstruction operator and a numerical flux function that incorporates the essential elements of LES, filtering and deconvolution. The optimization procedure starts from a nonlinearly stable numerical scheme and towards a final ALDM scheme which acts as an accurate sub-grid scale model. The viscous flux is discretized by a second-order central difference scheme and an explicit third-order total variation diminishing (TVD) Runge-Kutta scheme (Gottlieb & Shu 1998) is used for time marching. This method has been successfully applied to various supersonic flow cases, including shock wave/boundary layer interactions (SWBLI) on a flat plate (Pasquariello *et al.* 2017) and compression ramp (Grilli *et al.* 2012, 2013), and transition between regular and irregular shock patterns in SWBLI (Matheis & Hickel 2015), as well as in our previous work on SWBLI in laminar and transitional BFS flows (Hu *et al.* 2019, 2020). More details about the numerical method can be found in the literature (Hickel *et al.* 2006, 2014).

The numerical grids are generated using a Cartesian grid structure with block-based local refinement, as displayed in Figure 3. In addition, hyperbolic grid stretching was used in the wall-normal direction downstream of the step. The mesh is sufficiently refined near all walls with $y^+ < 0.9$ to ensure a well-resolved wall shear stress. The grid spacing becomes coarser with increasing wall distance but the expansion ratio between the adjacent blocks is not larger than two. The distribution of mesh cells are uniform in the spanwise direction. Using this discretization strategy, the computation domain has around $36 \times 10^6$ grid points and a spatial resolution of the flow field with $\Delta x^+_{max} \times \Delta y^+_{max} \times \Delta z^+_{max} = 36 \times 0.9 \times 18$ in wall units for the entire domain ($\Delta x^+_{max} = 0.9$ on the step wall). The temporal resolution, that is the time step, is approximately $\Delta t u_\infty / \delta_0 = 7.6 \times 10^{-4}$, corresponding to a Courant–Friedrichs–Lewy condition CFL $\leqslant 0.5$.

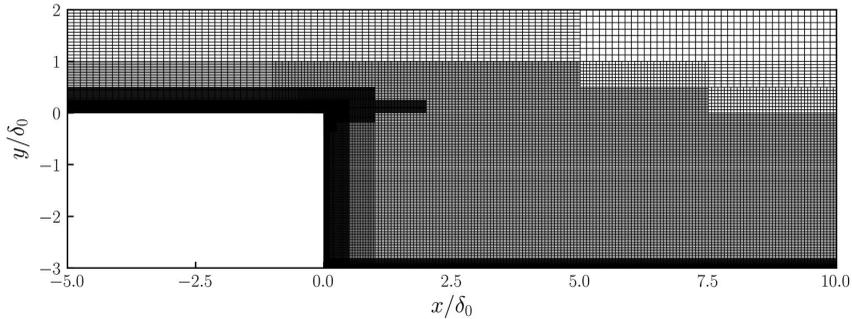

Figure 3: Details of the numerical grid in a x-y plane near the step. For clarity, the figures shows only every 2nd line in the $x$ direction and every 4th line in the $y$ direction.

The step and wall are modelled as no-slip adiabatic surfaces. All the flow variables are extrapolated at the outlet of the domain. On the top of the domain, non-reflecting boundary conditions based on Riemann invariants are used. Periodic boundary conditions are imposed in the spanwise direction. Inlet turbulent boundary conditions require a special approach since a very large domain for the natural development of turbulence is undesirable in view of computational resources and time. We use a synthetic turbulence generation method based on digital filter technique (Klein *et al.* 2003) to produce the appropriate turbulent inflow. This method can reproduce both first- and second-order statistical moments and spectra, without introducing low-frequency content which may modulate the low-frequency dynamics downstream. The reference data used are from Petrache *et al.* (2011) to specify realistic integral length scales and mean boundary layer profiles. According to previous studies (Grilli *et al.* 2013; Wang *et al.* 2015), a transient length of around $10\delta_0$ is sufficient for turbulence to develop in the supersonic boundary layer under these conditions. Nevertheless, we place the inflow plane $40\delta_0$ upstream of the step.

In order to examine the grid and domain size independence, the van Driest transformed mean velocity profile and Reynolds stresses in Morkovin scaling are provided at $x/\delta_0 = -5.0$ in Figure 4. The computed flow field reached a fully developed statistically steady state after an initial transient period of $t u_\infty / \delta_0 = 800$. The samples then were collected every $t u_\infty / \delta_0 = 0.5$ over an interval of another $t u_\infty / \delta_0 = 600$, yielding an ensemble size of 1200. For comparison, the figure also includes the theoretical law of the wall and incompressible DNS data of Schlatter & Örlü (2010) at $Re_\tau = 360$ and $Re_\theta = 1000$. The present mean velocity profile is consistent with both the logarithmic law of the wall (with the constants $\kappa = 0.41$ and $C = 5.2$) and the DNS data. The Reynolds stresses from the current LES are also in a good agreement with the reference data. Since the current LES data is for a compressible boundary layer that has a higher momentum thickness Reynolds number $Re_\theta = 2000$ and friction Reynolds number $Re_\tau = 400$, the velocity profile has a slight larger plateau value and streamwise Reynolds stress profile features with a higher peak value in the buffer layer (Marxen & Zaki 2019).

## 3. Results

### 3.1. *Mean flow features*

Figure 5 provides an overall view of the main flow topology. The upstream turbulent flow separates at the step edge and undergoes a centred Prandtl-Meyer expansion. The deflected shear layer travels downstream and finally reattaches on the downstream wall at $x/\delta_0 =$ 8.9. Compression waves are produced around the reattachment point, which coalesce into a reattachment shock oriented at an angle of 21° to the positive streamwise direction. Compared to the ramp and incident shock cases (Bonne *et al.* 2019; Priebe & Martín 2012), the freestream variables behind the interaction recover almost to their initial levels in the BFS configuration because there is only the weak reattachment shock generated by the compression waves whereas there are at least two stronger shocks in the other cases. The mean flow features of the laminar case are very similar to the present turbulent one, but the separated flow reattaches later at $x/\delta = 10.9$ and the mean shock angle is smaller, around 19° (Hu *et al.* 2019). These differences are caused by the stronger mixing in the turbulent case and are qualitatively consistent with existing experimental work (Zhi *et al.* 2014).

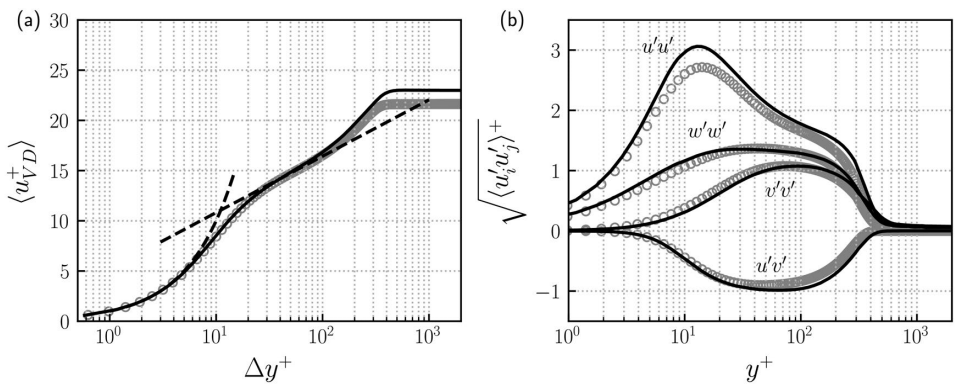

Figure 4: Mean profiles of the upstream turbulent boundary layer in inner scaling at $x/\delta_0 = -5.0$ with $Re_\tau = 400$ and $Re_\theta = 2000$. (a) Van Driest transformed mean velocity profile and (b) Reynolds stresses normalized by $\sqrt{\rho/\rho_w}$. ......, law of the wall; ———, present LES; ○, incompressible DNS data of Schlatter & Örlü (2010) at $Re_\tau = 360$ and $Re_\theta = 1000$.

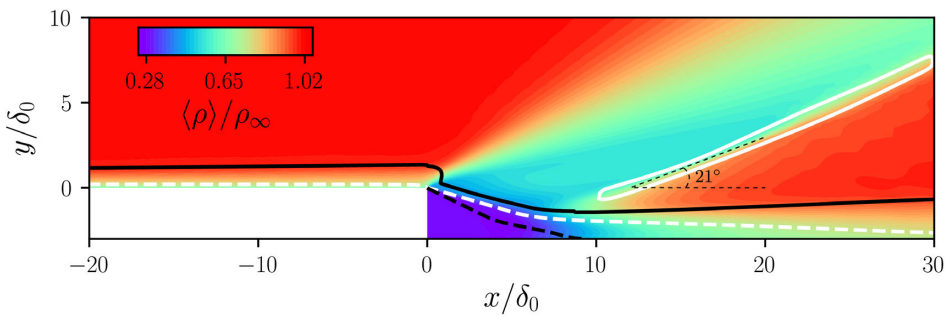

Figure 5: Density contours of the time- and spanwise-average flow field. The white dashed and solid lines denote the isolines of $Ma = 1.0$ and $|\nabla p|\delta_0/p_\infty = 0.24$. The black dashed and solid lines signify isolines of $u = 0.0$ and $u/u_e = 0.99$.

The mean reattachment length (equal to $L_r = x_r = 8.9\delta_0 \approx 3.0h$) is defined by the location of zero mean skin friction, $\langle C_f \rangle = 0$, in Figure 6(a). The value of $\langle C_f \rangle$ increases upstream of the step due to the flow acceleration induced by the expansion near the separation point ($x = 0$). Behind the step, there is a 'dead-air' zone where the recirculating velocity is extremely low. Thus, uniform $\langle C_f \rangle \approx 0$ are observed in the first 30% of the separation bubble ($0.0 \leqslant x/\delta_0 \leqslant 2.8$). The separated flow then rapidly reaches its strong level at $x \approx 2.1h \approx 6.2\delta_0$, which is very close to the value ($x \approx 2h \approx 6.4\delta_0$) reported by Chakravarthy *et al.* (2018). As the free shear layer reattaches on the downstream wall ($x/\delta_0 = 8.9$), the turbulent boundary layer recovers and $\langle C_f \rangle$ returns to a typical turbulent level ($\langle C_f \rangle = 0.0027$). The reattachment length $L_r \approx 3.0h$ based on the step height is in a good agreement with the previous experimental work by Bolgar *et al.* (2018) and the numerical study by Chakravarthy *et al.* (2018), who reported values of $L_r = 3.2h$ and $L_r = 3.0h$, respectively. Compared with the laminar case (blue dotted lines), the mean skin friction further confirms the shorter separation length in the turbulent case. The turbulent case has a much higher $\langle C_f \rangle$ upstream of the step. The laminar case reaches, however, a similar level downstream of the separation region, because laminar-to-turbulent transition is triggered within the separated shear layer.

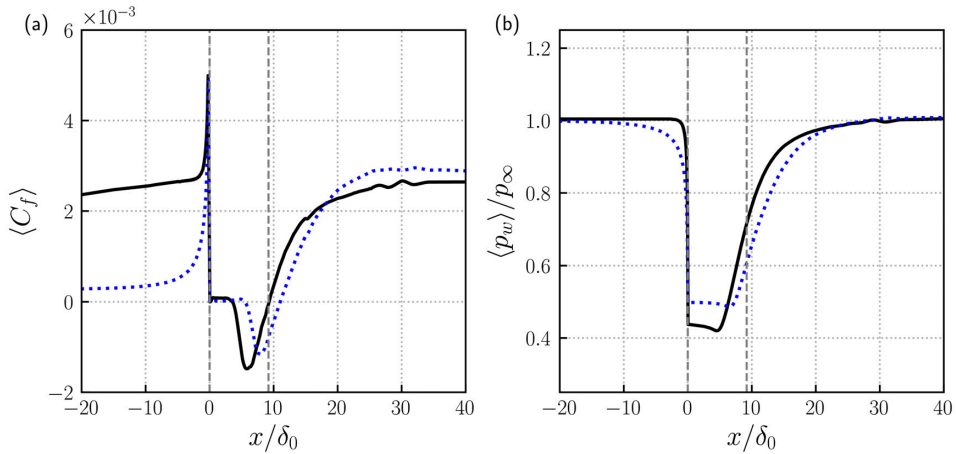

Figure 6: Streamwise variation of (a) skin friction and (b) wall pressure. The time- and spanwise-averaged values are indicated by the black solid lines (turbulent case) and blue dotted lines (laminar case). The vertical dashed line denotes the averaged separation and reattachment location for the turbulent case.

Figure 6(b) shows the streamwise variation of the wall pressure. As we can see, upstream the step, the wall pressure remains at almost the same level. The pressure drops drastically to around $42\% p_\infty$ in the first half of the separation bubble due to the expansion and the less energetic recirculating flow. The wall pressure then continues decreasing slowly to its global minimum at $x/\delta_0 = 4.6$, corresponding to the relatively strong reversed flow in terms of $\langle C_f \rangle$ in Figure 6(a). As the boundary layer reattaches on the wall and undergoes compression, the wall pressure quickly returns to the initial level. In the experimental work of Hartfield *et al.* (1993), they reported that the measured pressure deceases from 34.8 kPa to around 14.2 kPa ($\approx 41\% p_\infty$) upstream of the seperation bubble and returns to the free stream level downstream the interaction region, which are in a good agreement with the current results. In the laminar regime, the expansion fan is not as strong as for the turbulent case. Similarly, the intensity of the reattachment shock is weaker in the laminar case corresponding to a slower wall-pressure rise downstream.

### 3.2. *Instantaneous flow organisation*

Figure 7 visualizes the vortical structures using the $\lambda_2$ vortex criterion (Jeong & Hussain 1995). We see the expected small-scale coherent structures in the incoming turbulent boundary layer. Since the separated shear layer is inviscidly unstable, it rolls up and then larger and stronger vortical structures are generated over the bubble region. As the shear layer evolves downstream, the upstream small turbulent structures develop into larger coherent structures due to the shear layer instability, indicated by the arc-shaped vortices in the outer region of the boundary layer downstream the bubble. These coherent vortical structures propagate above the reversed flow from the separation to the reattachment location, and they also exist within the turbulent boundary layer downstream of the bubble.

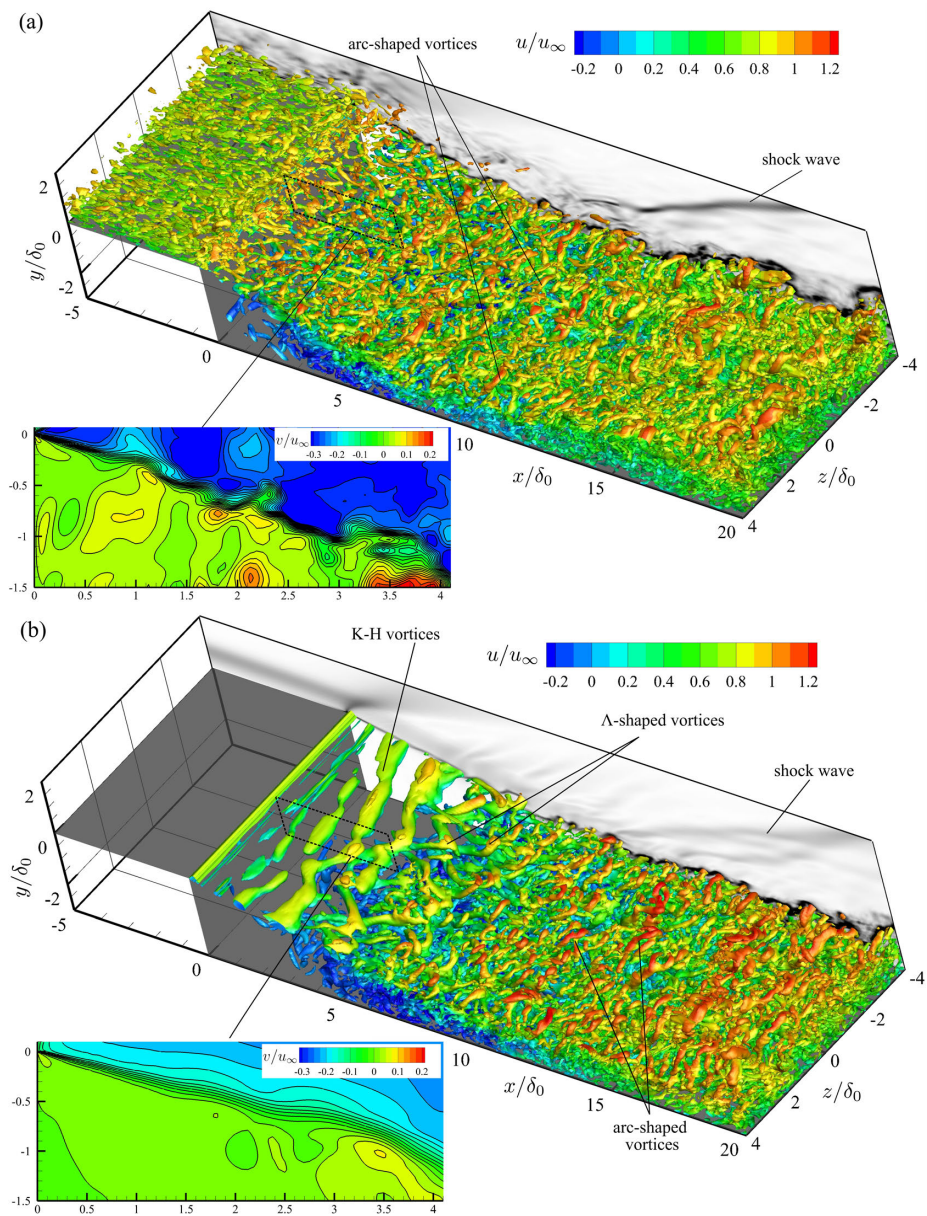

Figure 7: Instantaneous vortical structures at $tu_\infty/\delta_0 = 1000$, colored by contours of streamwise velocity, visualized by isosurfaces of $\lambda_2 = -0.08$. A numerical schlieren at $z/\delta_0 = -4.0$ slice is also included with $|\nabla\rho|/\rho_\infty = 0 \sim 1.4$. The roll-up structures in the shear layer are illustrated by contours of the wall-normal velocity at $z/\delta_0 = 0$ slice. (a) turbulent case and (b) laminar case.

For comparison, the instantaneous vortical structures of the laminar case are provided in Figure 7(b). The typical K-H vortex structure present in the laminar case is not observed in the current turbulent regime where the quasi two-dimensional vortices are probably distorted by the highly three-dimensional turbulence. In the middle of the shear layer, large coherent $\Lambda$-shaped vortices are formed and transformed into arc-shaped vortices downstream in the laminar case as a result of vortex stretching and tilting, whereas only arc-shaped vortices are present downstream in the turbulent case. From the numerical schlieren image shown on the $z/\delta_0 = -4$ slice, the shock intensity in the laminar case is weaker than that of the turbulent one, which is consistent with the evolution of the streamwise wall pressure in Figure 6(b).

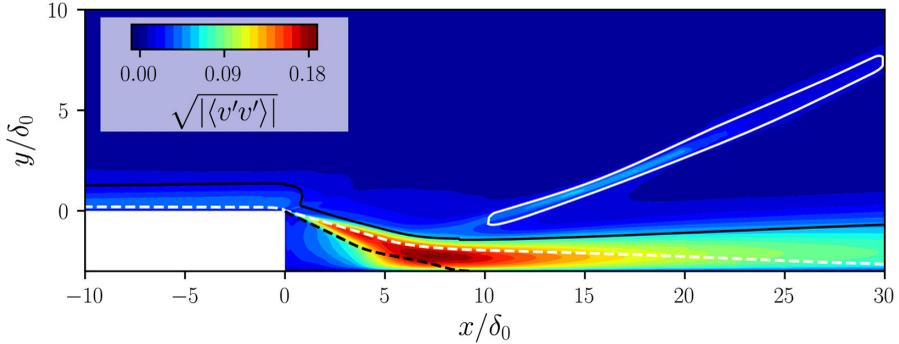

Figure 8: Contours of time- and spanwise-averaged variance of the wall-normal velocity. The white dashed and solid lines denote the isolines of $Ma = 1.0$ and $|\nabla p|\delta_0/p_\infty = 0.24$. The black dashed and solid lines signify isolines of $u = 0.0$ and $u/u_e = 0.99$.

### 3.3. *Unsteady characteristics*

The flow field over the BFS is highly unsteady, with vortices of various spatial scales observed in the visualization of Figure 7. To characterize the regions of most prominent unsteadiness, the variance of the wall-normal velocity is provided in Figure 8. As we can see, the most active region can be found along the separated shear layer (between the isoline of $u = 0$ and boundary layer edge), especially in the proximity of the reattachment location with a maximum of approximately $0.18u_\infty$ occurring at $x/\delta_0 = 7.2$, $y/\delta_0 = -2.2$. These major fluctuations caused by the recompression have also been reported in previous experimental work (Bolgar *et al.* 2018). Additionally, relatively weak fluctuations are found along the reattachment shock, reflecting its unsteady position. For the other normal Reynolds stress components $\langle u'u' \rangle$ and $\langle w'w' \rangle$, high levels of fluctuations are similarly observed around the reattachment point. We see that the separated shear layer and shock wave system is highly unsteady over the BFS with similar fluctuation intensities as in other canonical SWBLI geometries (Touber & Sandham 2011; Pasquariello *et al.* 2017).

Our attention then is put on the zones of the shear layer, reattachment location and shock wave to scrutinize the dynamic motions by examining a number of snapshots of the instantaneous flow field. First of all, we take a closer look at the shear layer. Figure 9 displays the contours of the streamwise velocity and streamlines at two arbitrarily selected instants. There are positive and negative streamwise velocity fluctuations alternating along the shear layer, which is the expected footprint of the shear layer instability behind the step. The convective Mach number $M_c$, defined as

$$M_c = \frac{u_1 - u_2}{a_1 + a_2} \quad , \tag{3.1}$$

is $M_c \approx 0.93$ at $x/\delta_0 = 4.5$, where $u_1$ and $u_2$ are the maximum streamwise velocity at the high-speed and low-speed sides of the mixing layer, and $a_1$, $a_2$ are the speed of sound at the corresponding locations. As indicated by Sandham & Reynolds (1991), the compressible shear layer also exhibit three-dimensional features at this convective Mach number, which explains the emergence of the three-dimensional waves in the shear layer behind the step shown in Figure 7. As the free shear flow evolves downstream, large coherent vortices are caused by the vortex pairing process behind the step (indicated by the streamlines of Figure 9), as reported by Soni *et al.* (2017). However, these shedding vortices are not typical two-dimensional structures in the turbulent regime, as we observe in Figure 7. For a BFS case, similar shedding vortices are observed both in subsonic and supersonic regimes (Tinney & Ukeiley 2009; Zhu *et al.* 2015).

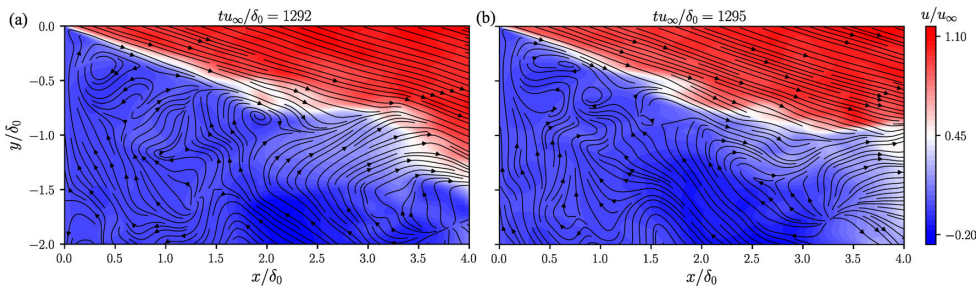

Figure 9: Contours of the instantaneous streamwise velocity for slice $z = 0$ at (a) $tu_\infty/\delta_0 = 1292$ and (b) $tu_\infty/\delta_0 = 1295$. The black arrow lines signify the streamlines.

Figure 10(a) shows the contours of the instantaneous skin friction coefficient. Distinctly different features are observed in the different regions of the flow field. In the upstream turbulent boundary layer, the levels of $C_f$ are homogeneously distributed and show clear evidence of the streamwise preferential orientation of the near-wall coherent structures. Figure 10(b) provides the spanwise wavenumber weighted power spectral density (PSD) of the streamwise wall velocity at two stations. As we can see, the wavenumber of the upstream structures ($x/\delta_0 = -5.0$) is $k_z \approx 2.0$, corresponding to a spanwise wavelength $\lambda_z \approx 0.5\delta_0$. The shear stress is relatively uniform at a low level downstream the step ($0 < x/\delta_0 < 5.0$) due to the less energetic flow in this region. Shortly upstream of the mean reattachment location ($5 < x/\delta_0 < 8.9$), there is significant reverse flow, cf. Figure 6(a), and $C_f$ indicates an increased spanwise length of the coherent structures. After reattachment, streamwise-oriented features are observed in the skin friction maps that indicate large scale streaks with a spanwise alternation of high and low velocity. For example at $x/\delta_0 = 10.0$, the dominant spanwise wavenumber of the streamwise wall velocity is $k_z \approx 0.35$ ($\lambda_z \approx 2.9\delta_0$), as shown in Figure 10(b). Further downstream, the intensity of $C_f$ becomes more homogeneous again. Similar phenomena have been reported in previous experiments of BFS with a wide range of Mach number (Ginoux 1971). The up-wash and down-wash effects of the Görtler-like vortices are believed to induce the alternating low and high skin friction in the spanwise direction around the reattachment, as will be discussed in the following sections. The characteristic wavelength of these streaks is between $\lambda_z = 2.0\delta_0$ and $3.3\delta_0$, which is consistent with previous experimental and numerical observations, reporting that the wavelength of these vortices is between two and three times the boundary layer thickness (Ginoux 1971; Priebe & Martín 2012; Grilli *et al.* 2013; Pasquariello *et al.* 2017).

In addition to the these relatively local phenomena, a large-scale unsteady motion is identified in the interaction system, as shown by the instantaneous velocity fields at two instants in Figure 11. These two instants represent different states of the separation bubble, i.e., expansion and shrinking. At $tu_\infty/\delta_0 = 954.5$, the length of separation bubble is around $L_r/\delta_0 = 7.5$, while the flow reattaches further downstream at about $x/\delta_0 = 9.0$ when expanding at $tu_\infty/\delta_0 = 1080$. In addition, the position of the shock (marked as white isolines of $|\nabla p|\delta_0/p_\infty = 0.4$) moves, most notably in the shock foot region. At $tu_\infty/\delta_0 = 954.5$, the shock foot locates somewhere between $x/\delta_0 = 7.5 \sim 10.0$ and the shock angle is $\eta = 22.2°$. At $tu_\infty/\delta_0 = 1080$, shock foot is between $x/\delta_0 = 5.0 \sim 7.5$ and shock angle reduces to $\eta = 16.8°$. It is clear from this comparison that the recirculation area and shock location vary in time.

For the laminar case, we also observe vortex shedding along the shear layer and the flapping motions of the shock (Hu *et al.* 2019). However, there are notable differences in the near wall dynamics, as can be seen when comparing the instantaneous skin friction contours and spanwise wavenumber weighted power spectral density in Figure 10 (turbulent case) to 12 (laminar case). The distribution of the laminar case skin friction is obviously spanwise uniform upstream the step. As the separated shear layer undergoes laminar-to-turbulent transition, the skin friction contours develop weak two-dimensional features around the reattachment location and further downstream. The dominant spanwise wavenumber of the upstream ($x/\delta_0 = -5.0$) and downstream ($x/\delta_0 = 10.0$) structures is close to each other with $k_z \approx 0.8$. The large low- and high-speed streaks are not observed downstream the reattachment point in the laminar case [Figure 12(a)] due to the small spanwise wavelength $\lambda_z \approx 1.2\delta_0$ ($\lambda_z \approx 2.9\delta_0$ around the reattachment in the turbulent case). This difference suggests that there is probably no evidence of the counter-rotating Görtler vortices in the laminar case.

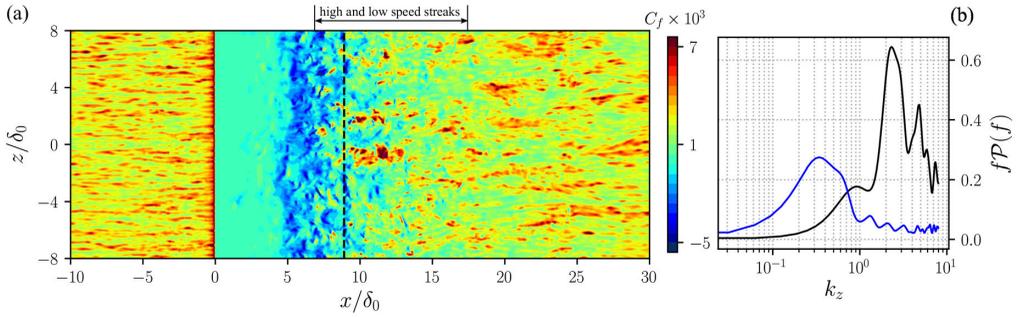

Figure 10: (a) contours of the instantaneous skin friction in the $x - z$ plane. The dashed line indicates the mean reattachment location. (b) spanwise wavenumber $k_z$ weighted power spectral density of the wall streamwise velocity (black line: $x/\delta_0 = -5.0$; blue line: $x/\delta_0 = 10.0$).

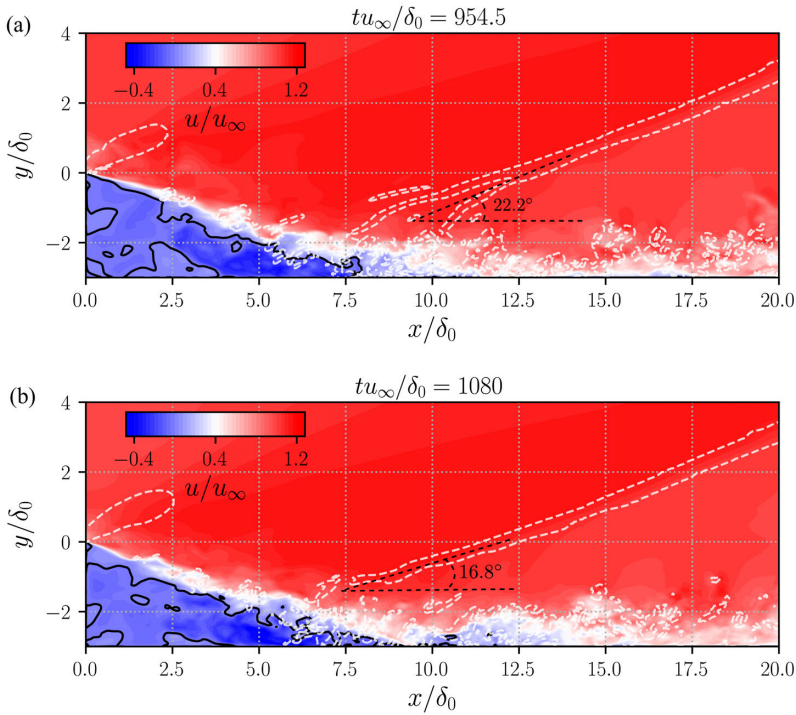

Figure 11: Contours of the instantaneous streamwise velocity for slice $z = 0$ at (a) $tu_\infty/\delta_0 = 954.5$ and (b) $tu_\infty/\delta_0 = 1080$. Black solid line denotes the isoline of $u = 0$ and white dashed line signifies the isoline of $|\nabla p|\delta_0/p_\infty = 0.4$.

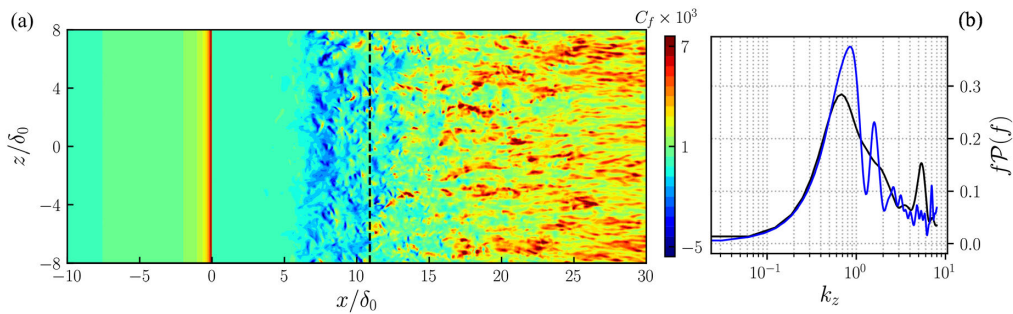

Figure 12: Laminar case: (a) contours of the instantaneous skin friction. The dashed line indicates the mean reattachment location. (b) spanwise wavenumber $k_z$ weighted power spectral density of the streamwise velocity (black line: $x/\delta_0 = -5.0$; blue line: $x/\delta_0 = 10.0$).

### 3.4. *Spectral Analysis*

An overview of frequency characteristics for the shock wave and separated boundary layer system is provided by the frequency weighted power spectral density of the wall pressure at selected streamwise locations in Figure 13. The sampling interval is $tu_\infty/\delta_0 = 950 \sim 1350$ with a sample frequency $f_s\delta_0/u_\infty = 4$, excluding the initial transient stage of the simulation. Welch's method with Hanning window was applied to compute the PSD using eight segments with 50% overlap (the same for the following PSD calculations). Upstream of the step ($x/\delta_0 = -3.0$), the spectrum shows a broadband bump centred around $St_\delta = f\delta_0/u_\infty = 0.8$, which is close to the characteristic frequency ($u_\infty/\delta$) of the upstream turbulent boundary layer (Dolling 2001). The low-frequency contents are relatively small upstream of the step, which demonstrates that the digital filter technique does not introduce significant spurious low-frequency features into the boundary layer. Downstream the step, we observe broadband low-frequency content between $St_\delta = 0.01 \sim 0.8$ ($St_h = fh/u_\infty = 0.03 \sim 2.4$), in addition to the typical signature of boundary layer turbulence at the higher frequencies. Two significant low frequencies can be identified along the streamwise distance. The lower one is around $St_\delta = 0.013$ (lower blue dashed line in the graph), which is most significant in a short distance behind the step ($x/\delta_0 \leqslant 3.0$). It appears that this low frequency is not the dominant one further downstream the separation bubble and an intermediate frequency at $St_\delta = 0.1 \sim 0.3$ (upper region separating by green dashed lines) begins to take the lead up to $x/\delta_0 = 20.0$. In the traditional ramp and impinging shock cases (Ganapathisubramani *et al.* 2007; Agostini *et al.* 2012; Pasquariello *et al.* 2017), the medium-frequency shear layer oscillations arise after the separation and the downstream propagation of this dynamics affects the reflected shock featuring intermediate frequencies, while the interaction between separation shock and boundary layer exhibits the low-frequency behaviour. Based on these conclusions, the medium frequency motions of the present BFS case are probably related to the shear layer instability, the downstream convection of which produces a relatively significant medium-frequency unsteadiness around the reattachment location ($x/\delta_0 = 9.25$). For the low-frequency contents of the BFS case, they are likely connected to the interactions of the reattachment shock and the separation bubble, the feedback of which leads to the low-frequency peak immediately downstream of the step ($x/\delta_0 = 1.0$).

To further confirm this conjecture, several aerodynamic parameters are extracted from the current results. For the medium-frequency behaviour, the temporal variation of the streamwise velocity within the separated shear layer and the spanwise-averaged reattachment position are plotted in Figure 14. These data are extracted with the same sampling frequency as the aforementioned pressure signal. The location of the spanwise-averaged reattachment point $x_r$ is obtained as follows: the isolines of the streamwise velocity $u = 0$ are collected at each time step; and in each spanwise plane the most downstream position meeting this condition ($u = 0$) is determined as the instantaneous value of $x_r$. An unsteady motion at a frequency around $St_\delta = 0.2$ ($St_h = 0.6$) appears energetically dominant for both shear layer velocity and reattachment location, which is more clear in the spectra of Figure 14. This medium frequency is the characteristic frequency of the shedding vortices within the shear layer. These vortices are shedding downstream as the shear layer and pass through the reattachment downstream the bubble, which explains that a similar frequency is observed in the spectrum of the reattachment location. There are also less energetic peaks at lower frequencies around $St_\delta = 0.03$, which will be discussed in the next paragraph. When taking a closer look on a short interval in Figure 15, the velocity signal of the shear layer is more periodic and regular. In contrast, the curve for the reattachment point follows a more sawtooth-like trajectory, along which its value undergoes a sharp drop when the reattachment point moves upstream, while it experiences a less rapid relaxation as the reattachment location shifts downstream, for instance around $tu_\infty/\delta_0 = 1160$. The sawtooth-like behaviour was also reported for incident shock and ramp cases (Priebe & Martín 2012; Pasquariello *et al.* 2017), and is attributed to the passage of shedding vortices formed in the shear layer near the reattachment.

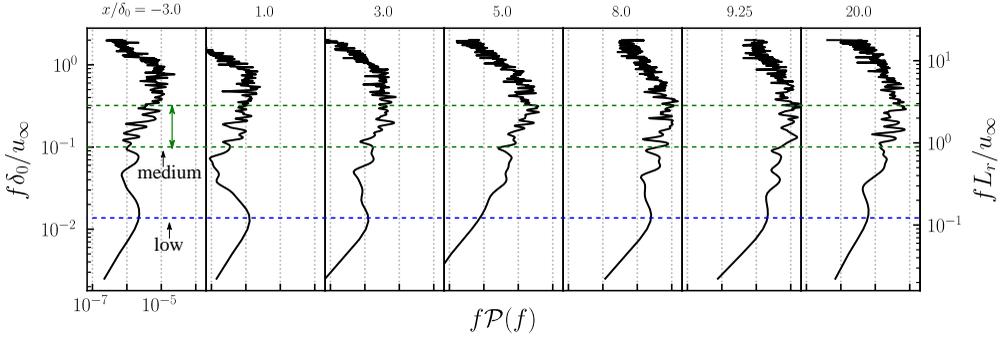

Figure 13: Frequency weighted power spectral density of the wall pressure with the streamwise distance.

With regard to the global dynamics, the temporal variation of the spanwise-averaged reattachment shock angle and separation bubble volume are shown in Figure 16. The bubble volume per unit spanwise length is defined as the area between the isoline of $u = 0$ and the bottom wall. The shock angle is determined based on the pressure gradient outside the boundary layer by fitting the isolines of $|\nabla p|\delta_0/p_\infty = 0.24$. We obtain two $x$ values by intersecting the isolines of $|\nabla p|\delta_0/p_\infty = 0.24$ at $y/\delta_0 = 0.5$ and then take the average of these two $x$ values as the first streamwise coordinate of the shock position. A second point of the shock position is obtained by repeating the same operation at $y/\delta_0 = 5.0$. A straight line is fitted based on these two points and the angle between the fitting line and the $x$−direction is considered as the shock angle. Both curves of the separation bubble size and shock angle are irregular and aperiodic in time, which suggests that the unsteady motion involves a range of time scales, cf. Refs (Dussauge *et al.* 2006; Priebe *et al.* 2016). For the signal of the separation bubble volume, shown in Figure 16(a), there is a significant low-frequency peak at $St_\delta = 0.023$ in the spectrum. It indicates that the bubble expands and shrinks with a frequency whose value is about two-order lower than the frequency of the typical turbulence. The spectrum of the shock angle also displays a peak at $St_\delta = 0.023$, see Figure 16(b), which is much more pronounced than the peak observed for the reattachment location at the same frequency in Figure 14(b). In addition, there is a second frequency peak around $St_\delta = 0.13$, which corresponds to the dominant frequency in the spectrum of reattachment location. Since the shock is formed by the compression waves originating at reattachment, spectra for the shock and reattachment locations include peaks at common frequencies.

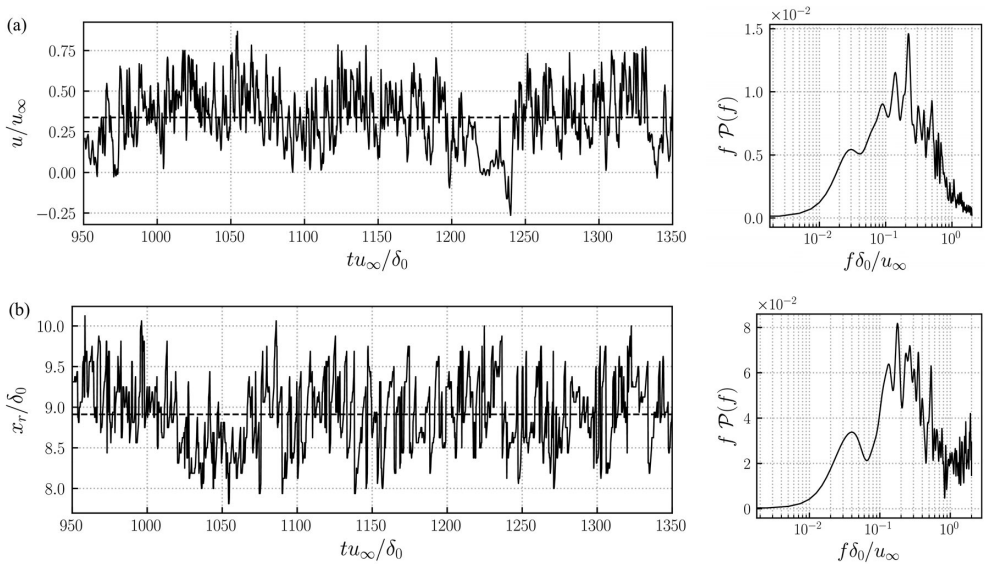

Figure 14: Temporal evolution and corresponding frequency weighted power spectral density of (a) streamwise velocity within the shear layer at $x/\delta_0 = 3.0625$, $y/\delta_0 = -1.0625$ and (b) the spanwise-averaged reattachment location. The black dashed line signifies the mean value.

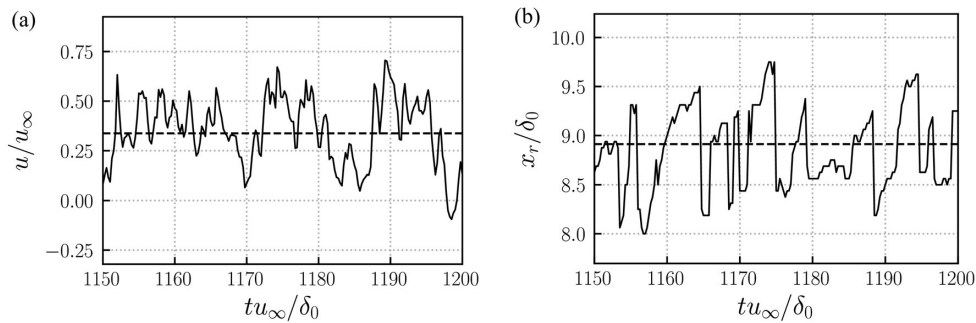

Figure 15: Details of figure 14 showing temporal evolution of (a) streamwise velocity within the shear layer at $x/\delta_0 = 3.0625$, $y/\delta_0 = -1.0625$ and (b) the spanwise-averaged reattachment location in a shorter period. The black dashed lines signify the mean values.

The statistical connection between the low-frequency signals can be quantified through coherence $C_{xy}$ and phase $\theta_{xy}$. The spectral coherence $C_{xy}$ between two temporal signals $x(t)$ and $y(t)$ is defined as

$$C_{xy}(f) = \left| P_{xy}(f) \right|^2 / \left( P_{xx}(f) P_{yy}(f) \right), 0 \leqslant C_{xy} \leqslant 1, \tag{3.2}$$

where $P_{xx}$ is the power spectral density of $x(t)$ and $P_{xy}(f)$ represents the cross-power spectral

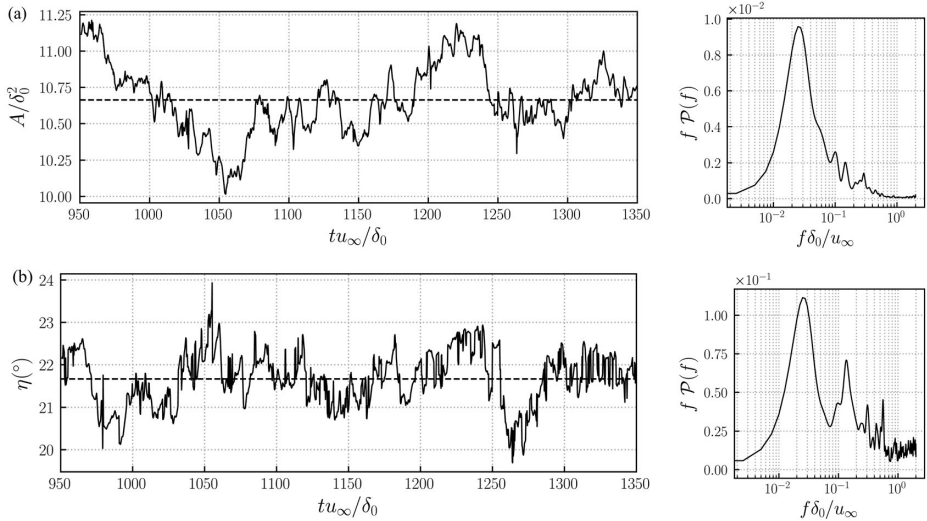

Figure 16: Temporal evolution and corresponding frequency weighted power spectral density of spanwise-averaged (a) bubble volume per unit spanwise length $A$ and (b) shock angle $\eta$. The black dashed line signifies the mean value.

density between signals $x(t)$ and $y(t)$. The phase $\theta_{xy}$ is determined by

$$\theta_{xy}(f) = \Im\left(P_{xy}(f)\right)/\Re\left(P_{xy}(f)\right), -\pi < \theta_{xy} \leqslant \pi. \tag{3.3}$$

For a specific frequency, if $0 < C_{xy} < 1$, it means that there is noise in the datasets or the relation between these two signals is not linear. When $C_{xy}$ equals to 1, it indicates that the signals $x(t)$ and $y(t)$ are linearly related, and $C_{xy} = 0$ signifies that they are completely unrelated.

The coherence and phase between the separation bubble volume and shock location of the spanwise-averaged snapshots are shown in Figure 17. The definition of the separation bubble volume is the same as before. The shock location is the $x$ coordinate of the intersection between $x$-axis and the fitted straight shock line (defined when calculating the shock angle in Figure 16). A high value of coherence ($C = 0.42$) is observed at the frequency $St_\delta = 0.028$, which manifests that the separation bubble and reattachment shock are nonlinearly related to each other around the shown dominant low frequency in the spectrum of Figure 16. Moreover, these two signals are approximately in phase, as can be seen from the low level of $\theta$. The above observations provide evidence that the unsteady low-frequency behaviour is related to the breathing of the separation bubble and the flapping motion of the shock, while the medium-frequency motions are associated with the shedding vortices of the shear layer. Thus a decoupling of the frequency scales is required to further trace the sustained source of the intrinsic unsteadiness of the interaction, which is the objective of §3.5.

Similar low- and medium-frequency are also observed for the laminar cases. Figure 18 plots the corresponding frequency weighted power spectral density of streamwise velocity around the mean reattachment location and the spanwise-averaged separation bubble size. To compare the laminar and turbulent cases, the frequency is rescaled by the reattachment length as $St_r = fL_r/u_\infty$. For the signal of streamwise velocity in Figure 18(a), the results show a broadband low-frequency spectrum for both the laminar and turbulence cases. However, a local spectrum peak is clearly observed at $St_r = 0.15$ ($St_\delta = 0.017$) in the turbulent case and at $St_r = 0.20$ ($St_\delta = 0.018$) for the laminar case. Since there are distinct shedding vortices in the shear layer for both flow regimes, the relevant prevailing medium frequencies are close to each other. For the bubble size in

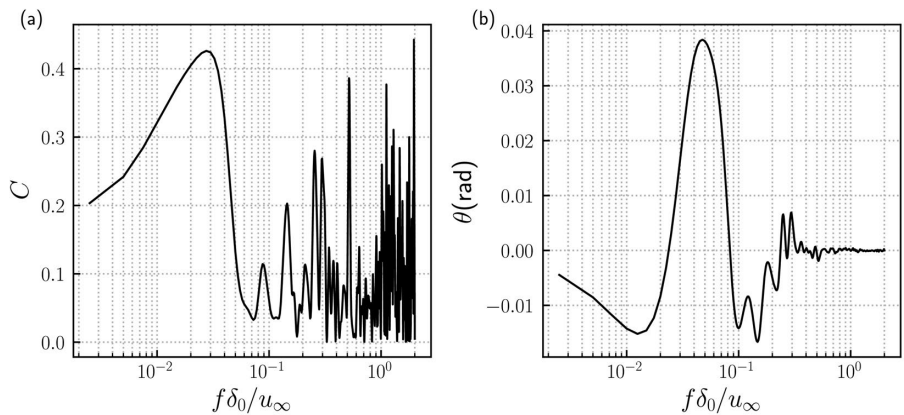

Figure 17: Statistical relation between the spanwise-averaged shock location and the volume of separation bubble per unit length: (a) coherence and (b) phase.

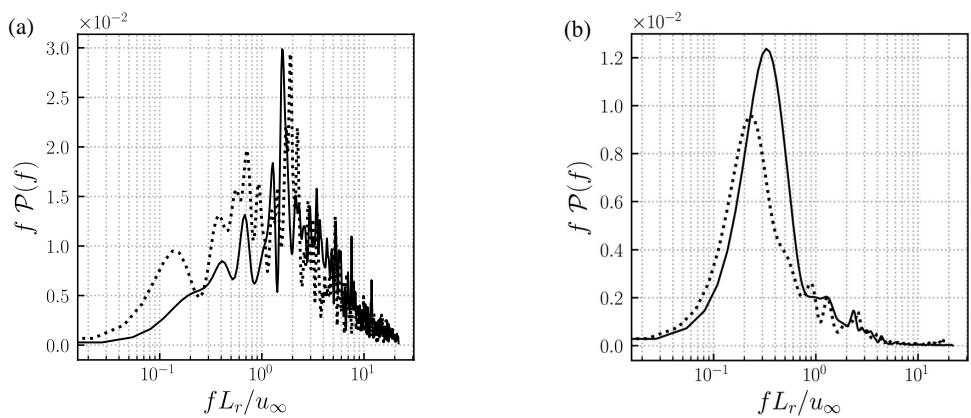

Figure 18: Frequency weighted power spectral density of (a) streamwise velocity around the mean reattachment location and (b) spanwise-averaged bubble volume per unit length $A$. The black solid line is the laminar case and dotted line represents the turbulent case.

Figure 18(b), the dominant frequency of the separation bubble in the laminar case is $St_r = 0.33$, while the corresponding value is lower ($St_r = 0.22$) in the turbulent case. These differences suggest that there are probably other flow dynamics involved, which leads to a lower frequency of the unsteady motions in the turbulent case. As previously discussed, Görtler-like vortices are likely to be associated with the low-frequency unsteadiness of SWBLI (Priebe *et al.* 2016). Therefore, we infer that the streamwise streaks in the turbulent regime may play a role in the transformation of the dominant low frequency.

### 3.5. *DMD analysis of the three-dimensional flow field*

To better separate the different dynamics from the coupled broadband frequency spectrum, a frequency-orthogonal modal decomposition of the three-dimensional flow field is conducted based on dynamic mode decomposition (DMD). Schmid (2010) first proposed this method to identify the most important dynamic information contained in equal-interval temporal snapshots from an unsteady flow field. Briefly summarized, a set of (reduced) modes will be extracted from the original dynamic system, each of which is associated with a single frequency behaviour and the combination of which approximate the complete unsteady system. Compared with proper orthogonal decomposition (POD), which is usually used for obtaining a low-dimensional reconstruction of the dynamic system, DMD focuses on the relevant flow dynamics while decoupling in frequency. This technique has been widely applied for various unsteady flow problems, such as the transition mechanism from laminar to turbulent flow (Hu *et al.* 2019), unsteadiness of the SWBLI (Pasquariello *et al.* 2017) and the identification of coherent vortex structures (Wang *et al.* 2020).

Following the DMD methodology, the original dynamic system can be represented by

$$\boldsymbol{Q}_1^{N-1} = \underbrace{[\phi_1, \phi_2, \cdots, \phi_{N-1}]}_{\phi} \underbrace{\begin{bmatrix} \alpha_1 & & & \\ & \alpha_2 & & \\ & & \ddots & \\ & & & \alpha_{N-1} \end{bmatrix}}_{\boldsymbol{D}_\alpha = \text{diag}\{\alpha\}} \underbrace{\begin{bmatrix} 1 & \mu_1 & \cdots & \mu_1^{N-1} \\ 1 & \mu_2 & \cdots & \mu_1^{N-1} \\ \vdots & \vdots & \ddots & \vdots \\ 1 & \mu_{N-1} & \cdots & \mu_{N-1}^{N-1} \end{bmatrix}}_{\boldsymbol{V}_{\text{and}}}. \quad (3.4)$$

where $\alpha_k$ can be considered as the amplitude of the $i$-th DMD mode $\phi_k$ and the Vandermonde matrix $\boldsymbol{V}_{\text{and}}$ signifies the temporal evolution of the dynamic modes. The eigenvalues $\mu_k$ are usually further converted into a more familiar complex stability plane through the logarithmic mapping $\lambda_k = \ln(\mu_k)/\Delta t$ (Leroux & Cordier 2016). The dynamic information about the growth rate $\beta_k$ and angular frequency $\omega_k$ of a specific DMD mode are then computed by

$$\begin{aligned} \beta_k &= \Re(\lambda_k) = \ln |\mu_k|/\Delta t \\ \omega_k &= \Im(\lambda_k) = \arctan(\mu_k)/\Delta t \end{aligned} \quad (3.5)$$

To facilitate a physical interpretation, we also reconstructed the real-valued flow field of the individual modes by superimposing the fluctuations from each mode $\phi_k$ onto the mean flow $q_m$, formulated as

$$q(x,t) = q_m + a_f \cdot \Re\left\{\alpha_k \phi_k e^{i\theta_k}\right\}, \theta_k = \omega_k t, \quad (3.6)$$

where $\alpha_k$ and $a_f$ are the amplitude and optional amplification factor of the corresponding mode $\phi_k$. The reconstructed flow field at different phase angles $\theta_k$ represents the temporal evolution of the dynamic system, In this way, the imaginary part of the reconstruction at a phase angle $\theta_i = 0$ is equivalent to the real part at $\theta_i = \pi$, and vice versa.

In the above analysis, we identified two types of unsteady behaviour at different frequencies. However, part of the signals were extracted from the spanwise-averaged field, like reattachment location, bubble size and shock angle; thus spanwise unsteady features may be missing from the two-dimensional flow field and a three-dimensional DMD analysis is required. Considering the large size of the data ensemble, a spatial subdomain $(-5.0 \leqslant x/\delta_0 \leqslant 20.0$ and $-3.0 \leqslant y/\delta_0 \leqslant 5.0$, covering the most interesting region) is extracted with a downsampled spatial resolution in all directions. The present DMD analysis of the three-dimensional subdomain is carried out based on 1200 equal-interval snapshots with the same temporal range of the previous signals and a smaller sampling frequency $f_s \delta_0/u_\infty = 2$ as the frequencies above the characteristic frequency of the turbulent integral scale $u_\infty/\delta_0$ are not of our current interest. The resulting frequency resolution is $1.67 \times 10^{-3} \leqslant St_\delta \leqslant 1$. Figure 19(a) shows the calculated eigenvalue spectrum provided by the standard DMD. The obtained modes appear as complex conjugate pairs and most of them are well distributed along the unit circle $|\mu_k| = 1$ except a few decaying modes within the circle, which means the resulting modes are saturated (Rowley *et al.* 2009). The magnitudes of the normalized amplitudes $(|\psi_k| = |\alpha_k|/|\alpha|_{\max})$ of the corresponding DMD modes are shown in Figure 19(b) for the positive frequencies and gray shaded by the growth rate $\beta_k$. Here, the strongly decaying modes $(|\mu_k| \leqslant 0.95)$ have been removed, as they do not contribute to the long-time flow evolution. The darker the vertical lines are, the less decayed the modes are. The convergence of the DMD results was verified by repeating the DMD using 400 snapshots less, which confirmed that the current DMD results are well-converged with respect to the number of the snapshots.

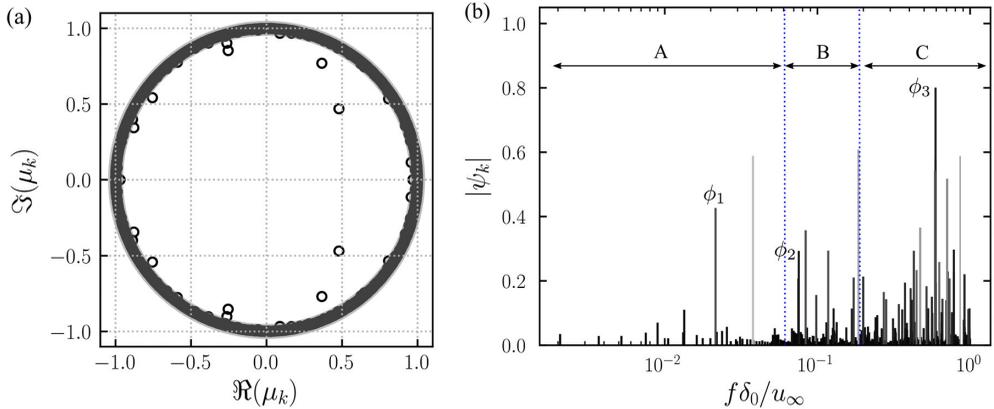

Figure 19: (a) Eigenvalue spectrum from the standard DMD and (b) normalized magnitudes for DMD modes with positive frequency, coloured by the growth rate $\beta_k$.

| Mode | $St_\delta$ | $|\psi_k|$ | $\beta_k$ |
|---|---|---|---|
| $\phi_1$ | 0.02151 | 0.42644 | -0.026404 |
| $\phi_2$ | 0.07546 | 0.29303 | -0.007900 |
| $\phi_3$ | 0.59361 | 0.80066 | -0.009204 |

Table 2: Information of the selected modes

From the frequency-magnitude spectrum, we identified three interesting frequencies, a lower one (marked as A) with $St_\delta < 0.06$, a medium one (marked as B) with $0.06 \leqslant St_\delta \leqslant 0.2$, and a higher one (marked as C) with $St_\delta > 0.2$. Based on the growth rate and magnitudes of the modes, three modes are selected from the frequency spectrum, one representative for each of the frequency ranges, labelled as mode $\phi_1$, $\phi_2$ and $\phi_3$. Table 2 provides the frequency, magnitudes and growth rate of these modes. All these modes have comparatively large magnitudes with $|\psi_k| > 0.1$. At the same time, these modes are the relatively darker ones in Figure 19(b) with decay rate $|\beta_k| < 0.03$, which suggests that their effects play a role during the entire process.

For the branch with lower frequencies, mode $\phi_1$ has been selected to illustrate the flow dynamics. Figure 20 shows the real part of the selected mode $\phi_1$ with the isosurfaces of the pressure fluctuations at phase angle $\theta = 0$ and $7\pi/8$. At both instants, the key features of this mode from the pressure signals are the significant structures along the shock and compression waves caused by the reattachment. Additionally, the fluctuations around the shock and reattachment are three dimensional, and indicate a slight wrinkling of the shock. Comparing the modal fluctuations at these two phases, the sign switch between them describes the oscillation of the reattachment shock. Figure 21(a) provides the pressure fluctuations at the slice $z = 0$, in which the effect that mode $\phi_1$ has on shock and compression waves is more clear. Note that perturbations in the upstream turbulent boundary layer are too weak to be visible at the given levels ($|p'/p_\infty| = 0.01$) of isosurfaces and in the contours.

In Figure 22, the fluctuations of the streamwise velocity component from DMD mode $\phi_1$ are given. As we can see, large fluctuations are aligned with the streamwise direction with negative and positive values alternating in the spanwise direction. These longitudinal structures appear to start within the fore part of the free shear layer and extend beyond the reattachment location. Additionally, they are mainly located in the near-wall part of the boundary layer, as shown by the streamwise velocity fluctuations in Figure 21(b). We also superimpose the modal fluctuations onto the mean flow and plot the contours of streamwise velocity in the $x - z$ slice at $y/\delta_0 = -2.875$ in Figure 23. The high- and low-speed streaks are obvious in the contours and show very similar features as the contours of skin friction in Figure 10. Other low-frequency modes between $St_\delta = 0.008 \sim 0.05$ have been also examined, and they all share common structures with mode $\phi_1$. The pressure and velocity fluctuations of DMD mode $\phi_1$ suggest that the low-frequency flapping motion of the shock is coupled with streamwise-elongated structures in the interaction region.

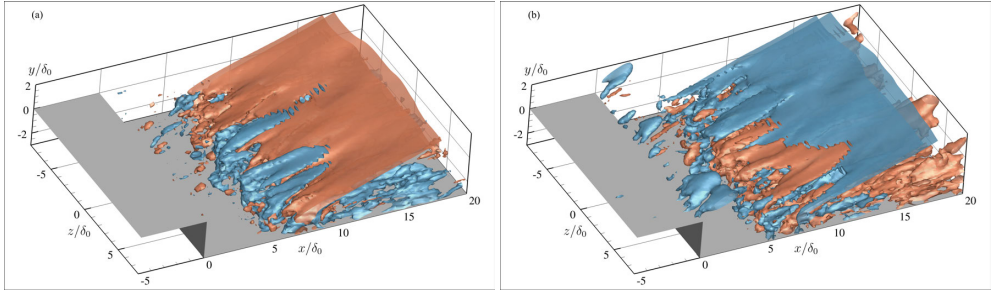

Figure 20: Isosurfaces of the pressure fluctuations from DMD mode $\phi_1$ with phase angle (a) $\theta = \pi/2$ and (b) $\theta = 7\pi/4$, only including the real part (red: $p'/p_\infty = 0.02$, blue: $p'/p_\infty = -0.02$).

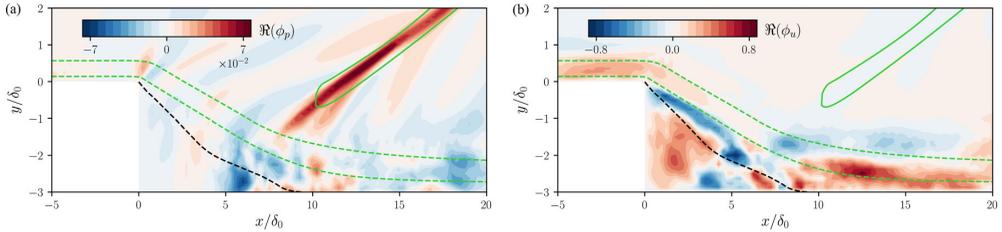

Figure 21: Real part of DMD mode $\phi_1$ indicating contours of modal (a) pressure fluctuations and (b) streamwise velocity fluctuations on the slice $z = 0$. The green solid line indicates the mean reattachment shock. The black dashed line signifies the dividing line. The green dashed lines represent the streamlines passing through $x/\delta_0 = 0$, $y/\delta_0 = 0.125$ and $x/\delta_0 = 0$, $y/\delta_0 = 0.5625$.

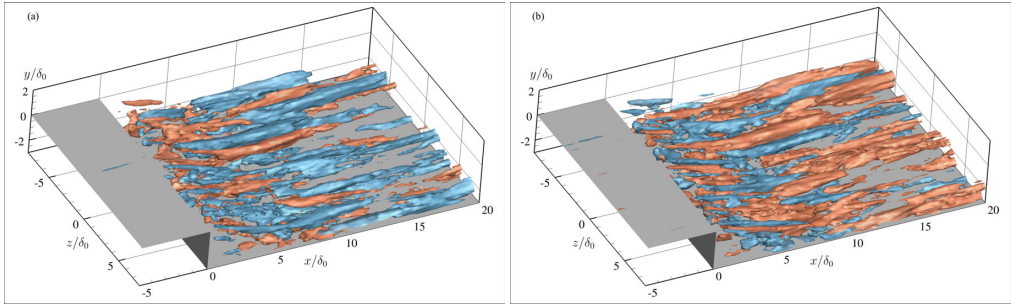

Figure 22: Isosurfaces of the streamwise velocity fluctuations from DMD mode $\phi_1$ with phase angle (a) $\theta = \pi/2$ and (b) $\theta = 7\pi/4$, only including the real part (red: $u'/u_\infty = 0.42$, blue: $u'/u_\infty = -0.42$).

These spanwise-aligned structures are the signature of counter-rotating Görtler-like vortices, as shown by the contours of modal streamwise vorticity in Figure 24. The location and strength of these vortex pairs are changing with the phase angles. At the given instants ($\theta = 0$ and $\theta = 5\pi/8$), the spanwise wavelength of the vortex pair is ranging from $1.9\delta_0$ to $1.6\delta_0$. To better visualize the unsteady motions represented by this mode, an animation of modal fluctuations with time is provided in the supplemental material (see supplementary movie 1). These time sequential snapshots of the reconstructed flow field from $\phi_1$ shows the flapping motion of the reattachment shock, i.e., the displacement of the shock around its mean shock location. The counter-rotating Görtler vortices are also unsteady in terms of their strength. Additionally, figure 24 indicates that these vortices can move in both the spanwise and wall-normal directions.

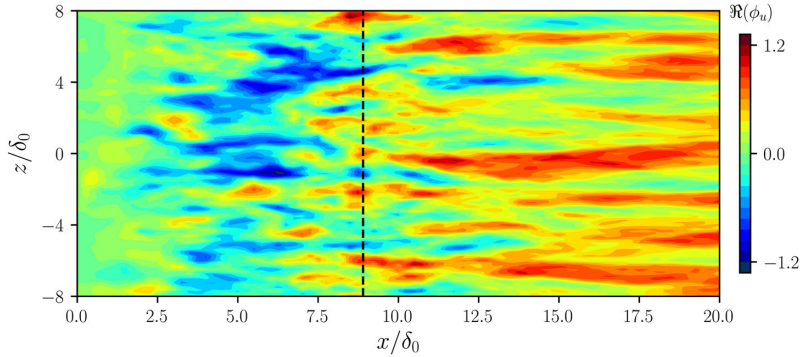

Figure 23: Contours of the reconstructed streamwise velocity from DMD mode $\phi_1$ in the $x - z$ plane at $y/\delta_0 = -2.875$.

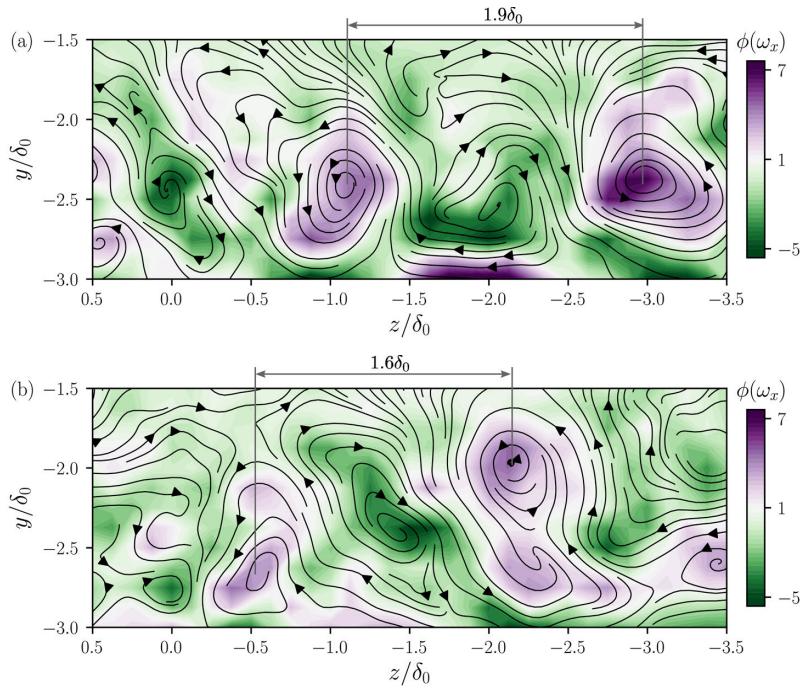

Figure 24: Contours of the streamwise vorticity from DMD mode $\phi_1$ with phase angle (a) $\theta = 0$ and (b) $\theta = 5\pi/8$ in the $z - y$ plane at $x/\delta_0 = 10$. Black arrow lines represent the streamlines on the slice.

For mode $\phi_2$, the pressure isosurfaces in Figure 25 show high levels of fluctuations along the reattachment shock, but the three-dimensional features are stronger compared to mode $\phi_1$. Positive and negative fluctuations are alternating in both spanwise and wall-normal directions, which represents a propagation of waves from the shear layer and outwards along the shock. The radiation of the Mach-like waves is in agreement with the results from a global linear stability analysis of an impinging shock case in a laminar regime (Guiho *et al.* 2016). The emission of these waves induces large disturbances along the streamwise direction in the supersonic part of the flow field. In the contours of modal spanwise-averaged pressure fluctuations in Figure 27, the radiation of the waves along the streamwise direction and shock is easier to observe.

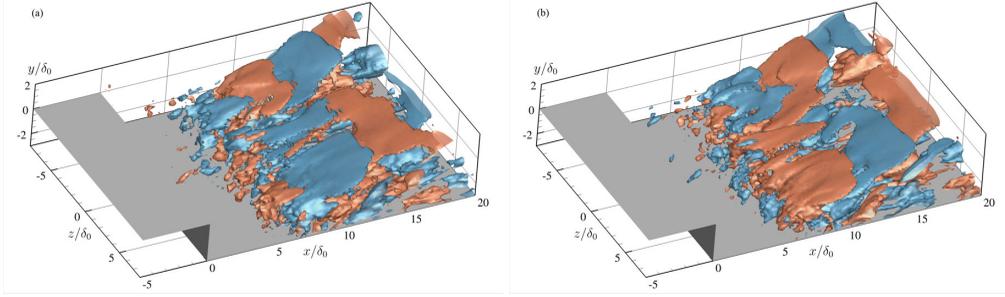

Figure 25: Isosurfaces of the pressure fluctuations from DMD mode $\phi_2$ with phase angle (a) $\theta = 0$ and (b) $\theta = 3\pi/4$, only including the real part (red: $p'/p_\infty = 0.02$, blue: $p'/p_\infty = -0.02$).

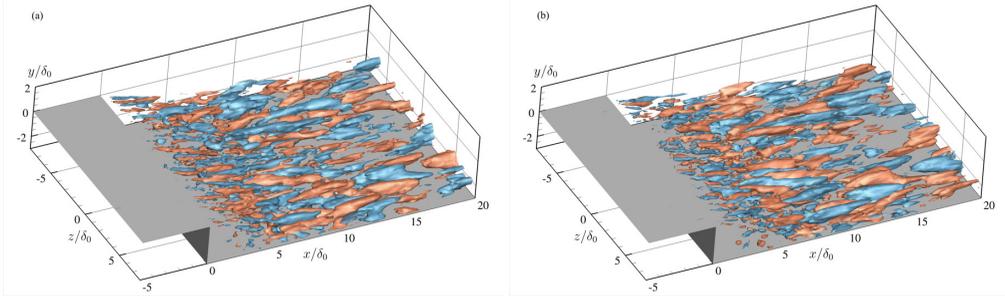

Figure 26: Isosurfaces of the streamwise velocity fluctuations from DMD mode $\phi_2$ with phase angle (a) $\theta = 0$ and (b) $\theta = 3\pi/4$, only including the real part (red: $u'/u_\infty = 0.3$, blue: $u'/u_\infty = -0.3$).

Considering the fluctuations of the streamwise velocity, shown in Figure 26, smaller longitudinal vortical structures are observed, compared to mode $\phi_1$. These vortices alternate along both the spanwise and streamwise directions, and are mainly concentrated within the boundary layer. Clearly, this mode represents the convection of the shear layer vortices and the induced Mach-like waves in the supersonic part of the flow field, which can also be seen in the contours of modal spanwise-averaged streamwise velocity in Figure 27(b). Similar observations were also reported in the two-dimensional DMD analysis of an incident shock case (Pasquariello *et al.* 2017). From the animation of the reconstructed flow based on this mode (see supplementary movie 2), the propagation of the Mach-like waves starting from the shock foot and the shedding of the small streamwise vortices are obvious.

The higher frequency modes, branch C, are related to the small-scale turbulent dynamics. For example for mode $\phi_3$, pressure fluctuations in Figure 28 show small highly three-dimensional arc-shaped vortices. These spanwise-aligned vortices are generated from the downstream part of the shear layer. The streamwise displacement of the fluctuations contours at different phase angles indicates the convection of the coherent vortices.

The convection behaviour of this mode is also evident from isosurfaces of the streamwise

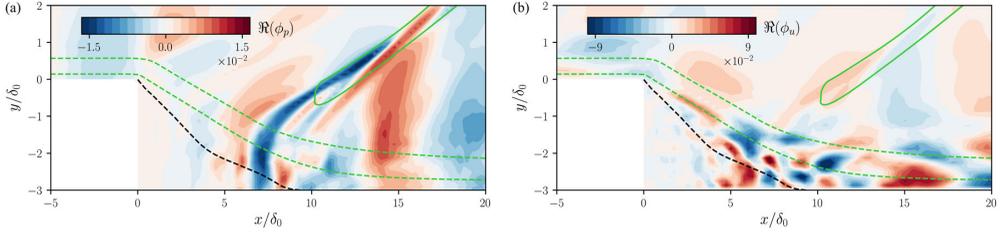

Figure 27: Real part of DMD mode $\phi_2$ indicating contours of modal spanwise-averaged (a) pressure fluctuations and (b) streamwise velocity fluctuations. The green solid line indicates the mean reattachment shock. The black dashed line signifies the dividing line. The green dashed lines represent the streamlines passing through $x/\delta_0 = 0$, $y/\delta_0 = 0.125$ and $x/\delta_0 = 0$, $y/\delta_0 = 0.5625$.

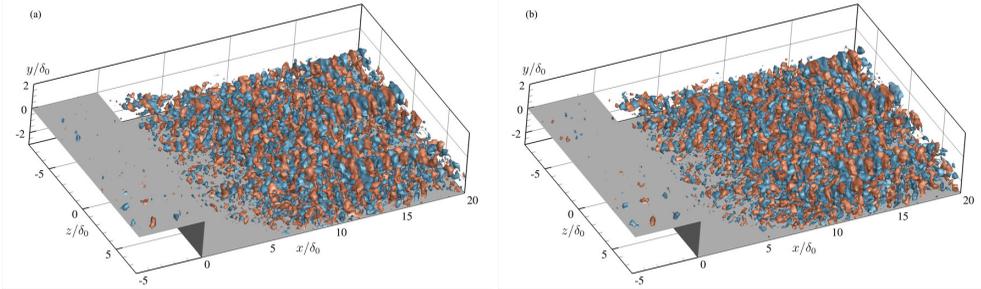

Figure 28: Isosurfaces of the pressure fluctuations from DMD mode $\phi_3$ with phase angle (a) $\theta = 0$ and (b) $\theta = 3\pi/8$, only including the real part (red: $p'/p_\infty = 0.06$, blue: $p'/p_\infty = -0.06$).

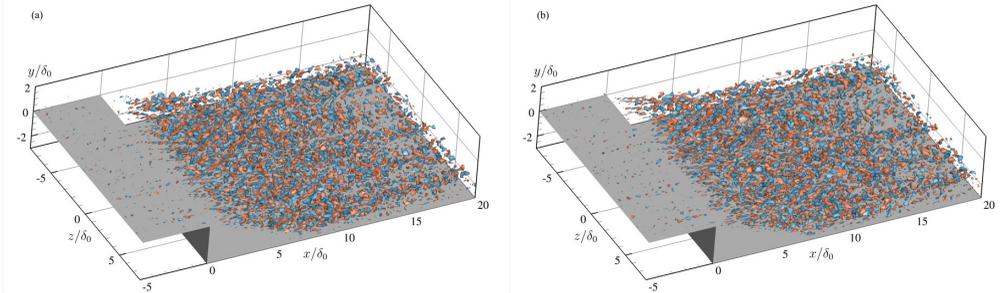

Figure 29: Isosurfaces of the streamwise velocity fluctuations from DMD mode $\phi_3$ with phase angle (a) $\theta = 0$ and (b) $\theta = 3\pi/8$ at slice $z = 0$, only including the real part (red: $u'/u_\infty = 0.6$, blue: $u'/u_\infty = -0.6$)

velocity fluctuations, shown in Figure 29. The velocity fluctuations originate from the strong shear layer behind the step. It is also noticed that these fluctuations are distributed along the free shear layer and downstream boundary layer. Additionally, this mode shows less anisotropic features, compared with the other two modes. The frequency of this mode is close to the typical frequency of the turbulence considering the thicker boundary layer downstream the step. Thus, we consider this mode to be related to the convection of typical turbulent structures (see supplementary movie 3) that result from an amplification of the incoming turbulence by the separation bubble, cf. the stability analysis of Guiho *et al.* (2016) for an incident shock SWBLI case.

We also performed a two-dimensional DMD analysis for the laminar case and similarly divided the modes into three branches (Hu *et al.* 2019). The branch with higher frequencies centred at $f\delta_0/u_\infty \approx 0.1$ is associated with the shedding of large coherent shear vortices, which is also observed in the present turbulent case. The other two branches with lower frequencies are related to the unsteady motions of the separation bubble and the shock. In contrast to the turbulent case, we found no evidence of Görtler-like vortices in the laminar case.

## 4. Physical mechanism of low-frequency unsteadiness

The current BFS case shows unsteady motions at similar low frequencies as those usually observed for SWBLI on flat plates and on compression ramps. Compared with these cases, however, the flow topology of the present case shows significantly different features. In canonical impinging shock and ramp cases, the separation bubble is enclosed by a separation shock and reattachment shock. In contrast, the recirculation region in a BFS case is surrounded by the step expansion fan, and reattachment shock. In terms of the mean skin friction, the recirculating flow is usually less uniform downstream of the mean separation position and recovers slower in the ramp and incident shock cases (Pasquariello *et al.* 2017; Priebe & Martín 2012). The fluctuations of $\langle C_f \rangle$ inside the separation bubble in these cases are usually related to the low-frequency unsteadiness. In the current case, however, the skin friction (Figure 6) is relatively uniform in the upstream part of the bubble, which is caused by the 'dead-air' region close to the stationary step wall. The wall pressure is usually increasing throughout the separation bubble in the ramp and incident shock cases (Priebe & Martín 2012; Sansica *et al.* 2016). In the current case, the pressure drastically drops at the step and keeps a relatively steady low level in the separation bubble, which is typical for strong interactions. These differences may suggest different low-frequency features between these cases. To compare with other canonical SWBLI cases, the dimensional frequencies are rescaled based on the separation length $L_r$ as $St_r = f L_r / u_\infty$ in the following discussion.

The instantaneous visualizations displayed in §3.3 illustrate both relatively localized and global unsteady motions in the flow field, involving high and low speed streaks, breathing bubble and oscillating reattachment shock, as well as vortex shedding in the separated shear layer. A linear stability analysis of the mean flow shows that the most unstable global mode is mainly distributed along the dividing line ($u = 0$), especially near the reattachment location (Guiho *et al.* 2016). The RMS wall-normal velocity in Figure 8 is consistent with this observation. The spectral analysis in section 3.4 reveals that there are two kinds of low-frequency unsteadiness in the interaction region and the lower frequency around $St_r = 0.18$ is associated with the coupling of separation bubble and reattachment shock wave. Furthermore, the three-dimensional DMD analysis separates the different dynamics contributing to low-frequency interaction. Apart from the unsteady separation bubble and reattachment shock, the low-frequency mode $\phi_1$ also reveals the unsteady Görtler-like vortices, see Figure 24. Although these Görtler vortices are rather weak compared to other energetic dynamics such that they are hard to identify in the vortical visualization in Figure 7, the skin friction contours in Figure 10 capture the footprint of the associated high- and low-speed streaks.

These observations share qualitative similarities to the low-frequency DMD modes calculated by Priebe *et al.* (2016). In their ramp case, the fluctuations of the low-frequency mode clearly show shocks (mainly separation shock) and longitudinal Görtler vortices near the reattachment. In the impinging shock case of Pasquariello *et al.* (2017), both the visualization of streamwise vorticity and the DMD analysis of the skin friction support the finding of the Görtler vortices downstream the reattachment location. Both report that the frequency of this unsteadiness is between $0.01 < St_r < 0.2$, while the current results for a BFS correspond to a Strouhal number range of $0.03 < St_r < 0.6$, which is three times larger than the values of other canonical cases.

We believe that the higher frequency in the current case is caused by the fixed separation location and confinement by the step wall. In the ramp and impinging shock cases, the recirculation region can move from both separation and reattachment sides. By comparison, the current case can only move in the reattachment part due to the limitation of the step, and it is reasonable to assume that this leads to a smaller oscillation amplitude and correspondingly to a higher frequency. This explanation is supported by the temporal evolution of the reattachment point. In Figure 15(b), the calculated minimum, mean and maximum reattachment location are $x_r/\delta_0 = 8.3, 8.9$ and 10.2, which leads to an oscillation range of about 15%$L_r$. For ramp cases, oscillations of up to 70%$L_r$ have been reported (Priebe & Martín 2012). Moreover, the separation length is around three times the maximum separation height ($L_r = 3h$, the maximum separation height equals to the step height) in our BFS case, whereas the recirculating flow regions are typically much thinner in ramp and impinging shock cases. Estruch-Samper & Chandola (2018) proposed an entrainment-recharge mechanism to associate the low-frequency unsteadiness with the shedding effects. In this theory, the Strouhal number of the low-frequency breathing can be related to the entrainment frequency by

$$St_r^{\text{low}} = \pi \alpha_\varepsilon \xi_B \delta'^2 \frac{L_r}{h} \left( X_r^{\text{ent}} \right)^2 St_r^{\text{ent}} \approx C_\varepsilon \frac{L_r}{h} \left( X_r^{\text{ent}} \right)^2 St_r^{\text{ent}}, \tag{4.1}$$

where $\alpha_\varepsilon$ is the length-to-thickness ratio of the shedding coherent structures; $\xi_B$ represents the percentage of the entrainment mass and $\delta'$ is the spreading rate of the mixing layer. Huang & Estruch-Samper (2018) showed that the variations of these three parameters between different cases are small if the incoming flow conditions are close, which results in an approximate constant $C_\varepsilon \approx 0.08$. The ratio of the bubble length to bubble height $L_r/h$ and the dimensionless entrainment length $X_r^{\text{ent}}$ depend on the specific geometry. In a similar entrainment and injection model by Piponniau et al. (2009), they consider the entrainment usually only occurs in the rear half of the separation bubble, i.e., the downward part of the shear layer, which leads to a dimensionless entrainment length $X_r^{\text{ent}} \approx 0.5$ in the impinging shock and ramp cases. The geometry dependent transformation factor is $C^{\text{ent}} = \left( X_r^{\text{ent}} \right)^2 L_r/h \approx 1.5$ for these canonical cases. In the current BFS case, the entrainment length $X_r^{\text{ent}}$ is one, which gives the transform factor $C^{\text{ent}} \approx 3$. The entrainment frequencies $St_r^{\text{ent}}$ of the shear layer shedding behaviour are similar for all these cases; thus the BFS case will yield a about two times larger $St_r^{\text{low}}$ than the impinging shock and ramp cases. However, this model only provides an estimate of the low frequency, and we do not expect to obtain an accurate value.

Several works in the literature have found evidence of Görtler-like vortices in SWBLI. Görtler vortices typically have a spanwise length-scale in the order of the incoming boundary layer thickness (Settles et al. 1979). The Görtler number, defined as,

$$G_t = \frac{\theta^{3/2}}{0.18 \delta^* |R|^{1/2}}, \tag{4.2}$$

gives an indication on whether such vortices can form (Smits & Dussauge 2006), where $R$ is the radius of curvature of the streamline, $\delta^*$ is the boundary layer displacement thickness and $\theta$ is the boundary layer momentum thickness. Figure 30 shows the curvature $\delta/R$ and Görtler number $G_t$ along two streamlines of the mean flow inside the shear layer (shown in Figure 21 and Figure 27). Streamline 1 is closer to the wall and passes through the coordinate $x/\delta_0 = 0$ and $y/\delta_0 = 0.125$. Significantly large streamline curvature occurs at the separation point and around the reattachment. Correspondingly, two distinct peaks are observed at these locations. The large curvature and Görtler number at the separation is mainly caused by the sudden change of the geometry at the step edge. In a laminar flow, the critical Görtler number is around $G_t = 0.6$ (marked as gray dot-dashed line) for a wide range of $Re$, above which the flow exhibits significant centrifugal instability and local Görtler vortices will emerge inside the boundary layer (Smits & Dussauge 2006). We see that the Görtler number is larger than the critical value between $7.7 \leqslant x/\delta_0 \leqslant 18.5$ and reaches its extremum $G_t = 1.2$ at $x/\delta_0 = 10.9$, close to the reattachment. Streamline 2 is in the middle of the boundary layer and passes through the point $x/\delta_0 = 0$ and $y/\delta_0 = 0.5625$. The spatial evolution of the curvature and Görtler number has a similar trend as for the streamline 1 and the Görtler number is above the critical value for $7.0 \leqslant x/\delta_0 \leqslant 19.3$.

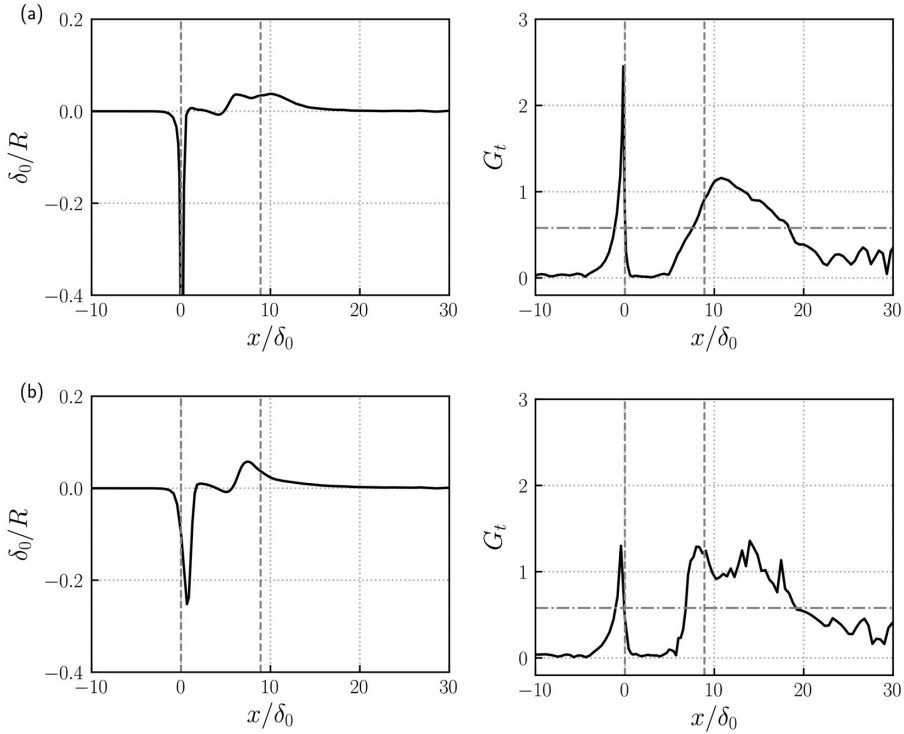

Figure 30: Curvature $\delta_0/R$ and Görtler number $G_t$ along two streamlines of the mean flow. (a) streamline 1 through $x/\delta_0 = 0$ and $y/\delta_0 = 0.125$, (b) streamline 2 through $x/\delta_0 = 0$ and $y/\delta_0 = 0.5625$. Vortical dashed lines indicate the separation and reattachment point. Horizontal dot-dashed line signify the critical $G_r$ in a laminar flow.

In Figure 10, we show that the high and low speed streaks are observed within the region $6.8 \leqslant x/\delta_0 \leqslant 17.8$, which is consistent with the zones with large Görtler number. The intrinsic spanwise wavelength of these streamwise vortex pairs is reported as $\lambda_z \approx 2\delta$ in SWBLI systems (Schülein & Trofimov 2011; Priebe et al. 2016), which is also in agreement with the current observations $\lambda_z = 2.0\delta_0 \sim 3.1\delta_0$. The streamwise velocity field reconstructed from DMD mode $\phi_1$ also displays these high and low speed streaks, see Figure 23. Although there is no general critical $G_t$ reported in the literature for turbulent separated flow, the high levels of $G_t$ provide an indication of sufficiently strong Görtler instability at the reattachment location, which provides an explanation for the low and high speed streaks observed in Figure 10 and the detected streamwise-oriented structures in DMD mode $\phi_1$.

Görtler vortices resulting from strong curvature could be unsteady in the turbulent flow, as concluded by Floryan (1991) from various experiments in low-speed turbulent flows. One of the situations proposed is that the generated streamwise-oriented vortices oscillate in the spanwise direction if they are affected by three-dimensional disturbances in the incoming flow. The unsteady streamwise vortices observed in the incident shock and ramp cases (Sun et al. 2012; Pasquariello et al. 2017; Priebe et al. 2016) both fall into this category. The mode structure shown in Figure 22 oscillates in spanwise direction. It is noticed that this proposition involves incoming disturbances, which may suggest a certain dependence on upstream flow conditions. From Figure 21(b), we can observe that weak upstream disturbances and fluctuations are part of the same DMD modes as the downstream Görtler vortices, which manifests that the observed Görtler vortices in the present study indeed have a significant correlation with upstream disturbances.

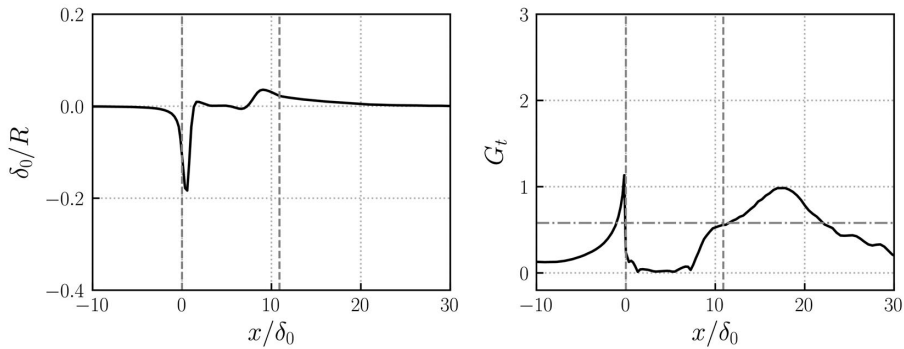

Figure 31: Curvature $\delta_0/R$ and Görtler number $G_t$ along the streamline through $x/\delta_0 = 0$ and $y/\delta_0 = 0.5625$ for the laminar case. Vortical dashed lines indicate the separation and reattachment point. Horizontal dot-dashed line signify the critical $G_r$ in a laminar flow.

For comparison, we also show the Görtler number for our laminar case (Hu *et al.* 2019) in Figure 31. As we can see, the curvature around the reattachment location in the laminar case is smaller than the one in the turbulent flow. As a result, $G_t$ has a smaller value than for the turbulent case (cf. Figure 30). Specifically, $G_t$ is below the critical value in the whole separation bubble, which probably is the reason that there are no Görtler vortices around the reattachment point in the laminar case (cf. Figure 12). In addition, this difference also shows how the existence of Görtler vortices is affected by upstream fluctuations: more turbulent incoming flow leads to a smaller separation length and thus stronger Görtler instability. On the other hand, the flow field around the reattachment location is more likely to reorganize and form into the spanwise-aligned vortices in the turbulent regime due to the incoming three-dimensional fluctuations.

Based on the above discussion, the following physical mechanism is proposed for the production of the low-frequency unsteadiness. The incoming turbulent flow experiences strong shear and curvature upon separation, which leads to large coherent vortical structures along the shear layer. Near the reattachment, there is significant centrifugal instability within the shear layer due to the concave streamlines. The Görtler instability, excited by incoming 3D turbulence, leads to large streamwise oriented vortices, which produce high- and low-speed streaks around the reattachment region (cf. Figure 10 and 23). These Görtler vortices are unsteady, which leads to spanwise shock wrinkling at very low frequencies, as we see from the streamwise velocity fluctuations from DMD mode $\phi_1$. Therefore, we believe that the centrifugal force and induced Görtler vortices are the main driving force of the global low-frequency unsteadiness in the turbulent case, which suggests that the low-frequency oscillation of SWBLI is inherently a three-dimensional mechanism. In the meantime, there is also notable dependence on the upstream fluctuations within the incoming turbulent boundary layer.

## 5. Conclusions

The unsteady dynamics of SWBLI over a BFS, with particular attention to the low-frequency unsteadiness, has been investigated at $Ma = 1.7$ using a well-resolved LES. The mean flow field illustrates the main flow topology of SWBLI in the BFS, consisting of a centred Prandtl-Meyer expansion fan originating from the fixed separation point, a separation bubble behind the step and a reattachment shock generating from the compression waves. Different from the canonical impinging shock and compression ramp cases, the separation point is stationary and only one shock occurs in the BFS case. The instantaneous flow field shows that the unsteady behaviour is however similar to other SWBLI configurations, including the vortex shedding in the separated shear layer, as well as the breathing of the separation bubble and a flapping shock motion. The spectral analysis shows that there is a broad band of low-frequency oscillations, which we classify into two branches with the dominant frequencies centred near $St_\delta = 0.02$ and 0.2 in the current case. The lower frequency dynamics is related to the unsteady separation bubble size, as well as the shock angle and position, while the second one connects to the shedding of shear layer vortices, which also affects the reattachment location.

Three-dimensional DMD analysis was used to reveal the characteristic mode structures that contribute to the observed unsteady behaviour. The low-frequency mode $\phi_1$ provides evidence for the statistical link between the shock motion (by pressure fluctuations) and the unsteady Görtler vortices (by the streamwise velocity fluctuations) around the reattachment position. The high- and low-speed streaks in the contours of $C_f$ in Figure 10 and the reconstructed velocity field in Figure 23 are the signature of these spanwise-aligned vortices. The medium-frequency mode $\phi_2$ represents shear-layer vortices and Mach-like waves. We thus believe that the unsteady Görtler vortices around the reattachment provide the unsteady forcing that sustains the low-frequency motions of shock and separation bubble. In particular through the comparison with a laminar inflow case (Hu *et al.* 2019), we show that the upstream fluctuations have a notable effect on the formation and existence of the unsteady spanwise-aligned Görtler vortices.

## Acknowledgements

This work was carried out on the Dutch national e-infrastructure with the support of SURF Cooperative. This work is part of the research programme 'Dynamics of a Backward/Forward-Facing Step in a Supersonic Flow' with project number 2019.045, which is partly financed by the Dutch Research Council (NWO).

## Declaration of Interests

The authors report no conflict of interest.

## REFERENCES

Agostini, Lionel, Larchevêque, Lionel, Dupont, Pierre, Debiève, Jean-François François & Dussauge, Jean-Paul 2012 Zones of influence and shock motion in a shock/boundary-layer interaction. *AIAA Journal* **50** (6), 1377–1387.

Andreopoulos, J. & Muck, K. C. 1987 Some new aspects of the shock-wave/boundary-layer interaction in compression-ramp flows. *Journal of Fluid Mechanics* **180**, 405–428.

Beresh, S. J., Clemens, N. T. & Dolling, D. S. 2002 Relationship between upstream turbulent boundary-layer velocity fluctuations and separation shock unsteadiness. *AIAA Journal* **40** (12), 2412–2422.

Bolgar, Istvan, Scharnowski, Sven & Kähler, Christian J. 2018 The Effect of the Mach Number on a Turbulent Backward-Facing Step Flow. *Flow Turbulence and Combustion* **101** (3), 653–680.

Bonne, N., Brion, V., Garnier, E., Bur, R., Molton, P., Sipp, D. & Jacquin, L. 2019 Analysis of the two-dimensional dynamics of a Mach 1.6 shock wave/transitional boundary layer interaction using a RANS based resolvent approach. *Journal of Fluid Mechanics* **862**, 1166–1202.

Chakravarthy, Kalyana, Arora, Konark & Chakraborty, Debasis 2018 Use of digitally filtered inflow conditions for LES of flows over backward facing steps. *European Journal of Mechanics, B/Fluids* **67**, 404–416.

Chen, Zhi, Yi, ShiHe, He, Lin, Tian, LiFeng & Zhu, YangZhu 2012 An experimental study on fine structures of supersonic laminar/turbulent flow over a backward-facing step based on NPLS. *Chinese Science Bulletin* **57** (6), 584–590.

Clemens, Noel T & Narayanaswamy, Venkateswaran 2014 Low-Frequency Unsteadiness of Shock Wave/Turbulent Boundary Layer Interactions. *Annual Review of Fluid Mechanics* **46** (1), 469–492.

Délery, Jean & Dussauge, Jean-Paul 2009 Some physical aspects of shock wave/boundary layer interactions. *Shock Waves* **19** (6), 453–468.

DOLLING, DAVID S 2001 Fifty Years of Shock-Wave/Boundary-Layer Interaction Research: What Next? *AIAA Journal* **39** (8), 1517–1531.

DUPONT, PIERRE, HADDAD, C & DEBIÈVE, J F 2006 Space and time organization in a shock-induced separated boundary layer. *Journal of Fluid Mechanics* **559** (2006), 255.

DUSSAUGE, JEAN-PAUL, DUPONT, PIERRE & DEBIÈVE, JEAN-FRANÇOIS 2006 Unsteadiness in shock wave boundary layer interactions with separation. *Aerospace Science and Technology* **10** (2), 85–91.

ERENGIL, MEHMET E. & DOLLING, DAVID S. 1991 Unsteady wave structure near separation in a Mach 5 compression rampinteraction. *AIAA Journal* **29** (5), 728–735.

ESTRUCH-SAMPER, DAVID & CHANDOLA, GAURAV 2018 Separated shear layer effect on shock-wave/turbulent-boundary-layer interaction unsteadiness. *Journal of Fluid Mechanics* **848**, 154–192.

FLORYAN, J.M. 1991 On the Görtler instability of boundary layers. *Progress in Aerospace Sciences* **28** (3), 235–271.

GAITONDE, DATTA V 2015 Progress in shock wave/boundary layer interactions. *Progress in Aerospace Sciences* **72**, 80–99.

GANAPATHISUBRAMANI, B., CLEMENS, N. T. & DOLLING, D. S. 2007 Effects of upstream boundary layer on the unsteadiness of shock-induced separation. *Journal of Fluid Mechanics* **585** (2007), 369–394.

GINOUX, JEAN J. 1971 Streamwise vortices in reattaching high-speed flows - A suggested approach. *AIAA Journal* **9** (4), 759–760.

GOTTLIEB, SIGAL & SHU, CHI-WANG 1998 Total variation diminishing Runge-Kutta schemes. *Mathematics of Computation* **67** (221), 73–85.

GREEN, J.E. 1970 Interactions between shock waves and turbulent boundary layers. *Progress in Aerospace Sciences* **11**, 235–340.

GRILLI, MUZIO, HICKEL, STEFAN & ADAMS, NIKOLAUS A 2013 Large-eddy simulation of a supersonic turbulent boundary layer over a compression–expansion ramp. *International Journal of Heat and Fluid Flow* **42**, 79–93.

GRILLI, MUZIO, SCHMID, PETER J., HICKEL, STEFAN & ADAMS, NIKOLAUS A. 2012 Analysis of unsteady behaviour in shockwave turbulent boundary layer interaction. *Journal of Fluid Mechanics* **700**, 16–28.

GUIHO, F, ALIZARD, F & ROBINET, J.-CH. 2016 Instabilities in oblique shock wave/laminar boundary-layer interactions. *Journal of Fluid Mechanics* **789**, 1–35.

HARTFIELD, ROY J., HOLLO, STEVEN D. & MCDANIEL, JAMES C. 1993 Planar measurement technique for compressible flows using laser-induced iodine fluorescence. *AIAA Journal* **31** (3), 483–490.

HICKEL, STEFAN, ADAMS, NIKOLAUS A & DOMARADZKI, J ANDRZEJ 2006 An adaptive local deconvolution method for implicit LES. *Journal of Computational Physics* **213** (1), 413–436.

HICKEL, STEFAN, EGERER, CHRISTIAN P & LARSSON, JOHAN 2014 Subgrid-scale modeling for implicit large eddy simulation of compressible flows and shock-turbulence interaction. *Physics of Fluids* **26** (10), 106101.

HU, WEIBO, HICKEL, STEFAN & VAN OUDHEUSDEN, BAS 2019 Dynamics of a supersonic transitional flow over a backward-facing step. *Physical Review Fluids* **4** (10), 103904.

HU, WEIBO, HICKEL, STEFAN & VAN OUDHEUSDEN, BAS 2020 Influence of upstream disturbances on the primary and secondary instabilities in a supersonic separated flow over a backward-facing step. *Physics of Fluids* **32** (5), 56102.

HUANG, XIN & ESTRUCH-SAMPER, DAVID 2018 Low-frequency unsteadiness of swept shock-wave/turbulent-boundary-layer interaction. *Journal of Fluid Mechanics* **856**, 797–821.

HUMBLE, R. A., ELSINGA, G. E., SCARANO, F. & VAN OUDHEUSDEN, B. W. 2009 Three-dimensional instantaneous structure of a shock wave/turbulent boundary layer interaction. *Journal of Fluid Mechanics* **622**, 33–62.

JEONG, JINHEE & HUSSAIN, FAZLE 1995 On the identification of a vortex. *Journal of Fluid Mechanics* **285**, 69–94.

KLEIN, M., SADIKI, A. & JANICKA, J. 2003 A digital filter based generation of inflow data for spatially developing direct numerical or large eddy simulations. *Journal of Computational Physics* **186** (2), 652–665.

LEROUX, ROMAIN & CORDIER, LAURENT 2016 Dynamic mode decomposition for non-uniformly sampled data. *Experiments in Fluids* **57** (5), 94.

LOTH, E., KAILASANATH, K. & LOHNER, R. 1992 Supersonic flow over an axisymmetric backward-facing step. *Journal of Spacecraft and Rockets* **29** (3), 352–359.

MARXEN, OLAF & ZAKI, TAMER A. 2019 Turbulence in intermittent transitional boundary layers and in turbulence spots. *Journal of Fluid Mechanics* **860**, 350–383.

MATHEIS, JAN & HICKEL, STEFAN 2015 On the transition between regular and irregular shock patterns of shock-wave/boundary-layer interactions. *Journal of Fluid Mechanics* **776**, 200–234.

MCCLURE, WILLIAM BERTON 1992 An experimental study of the driving mechanism and control of the unsteady shock induced turbulent separation in a Mach 5 compression corner flow. PhD thesis, University of Texas at Austin, Austin.

PASQUARIELLO, VITO, HICKEL, STEFAN & ADAMS, NIKOLAUS A. 2017 Unsteady effects of strong shock-wave/boundary-layer interaction at high Reynolds number. *Journal of Fluid Mechanics* **823**, 617–657.

PETRACHE, OANA, HICKEL, STEFAN & ADAMS, NIKOLAUS 2011 Large-Eddy Simulations of Turbulence Enhancement due to Forced Shock Motion in Shock-Boundary Layer Interaction. In *17th AIAA International Space Planes and Hypersonic Systems and Technologies Conference*, pp. 1–13. Reston, Virigina: American Institute of Aeronautics and Astronautics.

PIPONNIAU, S., DUSSAUGE, J. P., DEBIÈVE, J. F. & DUPONT, P. 2009 A simple model for low-frequency unsteadiness in shock-induced separation. *Journal of Fluid Mechanics* **629** (2009), 87–108.

PIROZZOLI, SERGIO & GRASSO, FRANCESCO 2006 Direct numerical simulation of impinging shock wave/turbulent boundary layer interaction at M=2.25. *Physics of Fluids Fluids* **18** (6), 065113.

PLOTKIN, KENNETH J. 1975 Shock wave oscillation driven by turbulent boundary-layer fluctuations. *AIAA Journal* **13** (8), 1036–1040.

POGGIE, J. & SMITS, A. J. 2001 Shock unsteadiness in a reattaching shear layer. *Journal of Fluid Mechanics* **429** (2001), 155–185.

PORTER, KEVIN M. & POGGIE, JONATHAN 2019 Selective upstream influence on the unsteadiness of a separated turbulent compression ramp flow. *Physics of Fluids* **31** (1).

PRIEBE, STEPHAN & MARTÍN, M. PINO 2012 Low-frequency unsteadiness in shock wave–turbulent boundary layer interaction. *Journal of Fluid Mechanics* **699** (2012), 1–49.

PRIEBE, STEPHAN, TU, JONATHAN H., ROWLEY, CLARENCE W. & MARTÍN, M. PINO 2016 Low-frequency dynamics in a shock-induced separated flow. *Journal of Fluid Mechanics* **807** (2016), 441–477.

ROWLEY, CLARENCE W., MEZIĆ, IGOR, BAGHERI, SHERVIN, SCHLATTER, PHILIPP & HENNINGSON, DAN S. 2009 Spectral analysis of nonlinear flows. *Journal of Fluid Mechanics* **641**, 115–127.

SANDHAM, N. D. & REYNOLDS, W. C. 1991 Three-dimensional simulations of large eddies in the compressible mixing layer. *Journal of Fluid Mechanics* **224**, 133–158.

SANSICA, ANDREA, SANDHAM, NEIL D. & HU, ZHIWEI 2016 Instability and low-frequency unsteadiness in a shock-induced laminar separation bubble. *Journal of Fluid Mechanics* **798**, 5–26.

SCHLATTER, PHILIPP & ÖRLÜ, RAMIS 2010 Assessment of direct numerical simulation data of turbulent boundary layers. *Journal of Fluid Mechanics* **659**, 116–126.

SCHMID, PETER J. 2010 Dynamic mode decomposition of numerical and experimental data. *Journal of Fluid Mechanics* **656** (2010), 5–28.

SCHÜLEIN, ERICH & TROFIMOV, VICTOR M. 2011 Steady longitudinal vortices in supersonic turbulent separated flows. *Journal of Fluid Mechanics* **672**, 451–476.

SETTLES, GARY S., FITZPATRICK, THOMAS J. & BOGDONOFF, SEYMOUR M. 1979 Detailed study of attached and separated compression corner flowfields in high reynolds number supersonic flow. *AIAA Journal* **17** (6), 579–585.

SMITS, ALEXANDER J & DUSSAUGE, JEAN PAUL 2006 *Turbulent shear layers in supersonic flow*. Springer Science & Business Media.

SONI, R. K., ARYA, N. & DE, ASHOKE 2017 Characterization of turbulent supersonic flow over a backward-facing step. *AIAA Journal* **55** (5), 1511–1529.

SOUVEREIN, LOUIS J., DUPONT, PIERRE, DEBIÈVE, JEAN-FRANCOIS, DUSSAUGE, JEAN-PAUL, VAN OUDHEUSDEN, BAS W. & SCARANO, FULVIO 2010 Effect of Interaction Strength on Unsteadiness in Shock-Wave-Induced Separations. *AIAA Journal* **48** (7), 1480–1493.

SRIRAM, A. T. & CHAKRABORTY, DEBASIS 2011 Numerical Exploration of Staged Transverse Injection into Confined Supersonic. *Defence Science Journal* **61** (1), 3–11.

SUN, Z., SCHRIJER, F. F.J., SCARANO, F. & VAN OUDHEUSDEN, B. W. 2012 The three-dimensional flow organization past a micro-ramp in a supersonic boundary layer. *Physics of Fluids* **24** (5), 055105.

THOMAS, F. O., PUTNAM, C. M. & CHU, H. C. 1994 On the mechanism of unsteady shock oscillation in shock wave/turbulent boundary layer interactions. *Experiments in Fluids* **18** (1-2), 69–81.

TINNEY, C. E. & UKEILEY, L. S. 2009 A study of a 3-D double backward-facing step. *Experiments in Fluids* **47** (3), 427–438.

TOUBER, EMILE & SANDHAM, NEIL D 2009 Large-eddy simulation of low-frequency unsteadiness in a turbulent shock-induced separation bubble. *Theoretical and Computational Fluid Dynamics* **23** (2), 79–107.

TOUBER, EMILE & SANDHAM, NEIL D. 2011 Low-order stochastic modelling of low-frequency motions in reflected shock-wave/boundary-layer interactions. *Journal of Fluid Mechanics* **671** (2011), 417–465.

UNALMIS, O. & DOLLING, D. 1996 On the possible relationship between low frequency unsteadiness of shock-induced separated flow and Goertler vortices. In *Fluid Dynamics Conference*, pp. 1–18. Reston, Virigina: American Institute of Aeronautics and Astronautics.

WANG, BO, SANDHAM, NEIL D., HU, ZHIWEI & LIU, WEIDONG 2015 Numerical study of oblique shock-wave/boundary-layer interaction considering sidewall effects. *Journal of Fluid Mechanics* **767**, 526–561.

WANG, JINCHUN, HUANG, GUOPING, LU, WEIYU & SULLIVAN, PIERRE E. 2020 Dynamic mode decomposition analysis of flow separation in a diffuser to inform flow control strategies. *Journal of Fluids Engineering* **142** (2).

WU, MINWEI & MARTÍN, M. PINO 2008 Analysis of shock motion in shockwave and turbulent boundary layer interaction using direct numerical simulation data. *Journal of Fluid Mechanics* **594**, 71–83.

ZHI, CHEN, SHIHE, YI, LIN, HE, YANGZHU, ZHU, YONG, GE & YU, WU 2014 Spatial density fluctuation of supersonic flow over a backward-facing step measured by nano-tracer planar laser scattering. *Journal of Visualization* **17** (4), 345–361.

ZHU, YANGZHU, YI, SHIHE, GANG, DUNDIAN & HE, LIN 2015 Visualisation on supersonic flow over backward-facing step with or without roughness. *Journal of Turbulence* **16** (7), 633–649.